\documentclass[fleqn,usenatbib]{mnras}

\usepackage{ae,aecompl}
\usepackage{graphicx}
\usepackage{amsmath,amsfonts,amssymb}
\usepackage[T1]{fontenc}

\newcommand{\nuc}[2]{${}^{#2} \rm #1$}

\def\gtaprx {\lower .14ex\hbox{\rlap{\raise .9ex\hbox{\hskip .3ex
	{\ifmmode{\scriptscriptstyle >}\else
		{$\scriptscriptstyle >$}\fi}}}
	\kern -.4ex{\ifmmode{\scriptscriptstyle \sim}\else
		{$\scriptscriptstyle\sim$}\fi}}}
\def\ltaprx {\lower .14ex\hbox{\rlap{\raise .9ex\hbox{\hskip .3ex
	{\ifmmode{\scriptscriptstyle <}\else
		{$\scriptscriptstyle <$}\fi}}}
	\kern -.4ex{\ifmmode{\scriptscriptstyle \sim}\else
		{$\scriptscriptstyle\sim$}\fi}}}

\newcommand{\s}{ \, {\rm s} }
\newcommand{\K}{ {\,\rm K} }

\newcommand{\erg}{\, {\rm erg} }

\newcommand{\km}{{\, \rm km}}
\newcommand{\ms}{{\, \rm ms}}

\newcommand{\nue}{\nu_{\rm e}} 
\newcommand{\nueb}{{\bar \nu}_{\rm e}} 
\newcommand{\nux}{\nu_x}

\newcommand{\del}[2]%
{\frac{\mathrm{d}{#2}}{\mathrm{d}{#1}}}
\newcommand{\Del}[2]%
{\frac{D{#2}}{D{#1}}}\newcommand{\ddel}[2]%
{\frac{\mathrm{d}^2{#2}}{\mathrm{d}{#1}^2}}

\newcommand{\pdel}[2]%
{\frac{\partial{#2}}{\partial{#1}}}
\newcommand{\pddel}[2]%
{\frac{\partial^2{#2}}{\partial{#1}^2}}

\newcommand{\Ms}{M_{\odot}}
\newcommand{\Mms}{M_{\rm ms}}

\title[Impact of asymmetric neutrinos on nucleosynthesis]
{The impact of asymmetric neutrino emissions on nucleosynthesis in core-collapse supernovae II -- progenitor dependences -- }

\author[S. Fujimoto and H. Nagakura]{Shin-ichiro Fujimoto$^{1}$
\thanks{E-mail: fuji@kumamoto-nct.ac.jp},
Hiroki Nagakura$^{2}$
\\
$^{1}$National Institute of Technology, Kumamoto College, Kumamoto 861-1102, Japan
\\
$^{2}$Department of Astrophysical Sciences, Princeton University, Princeton, NJ 08544, USA\\
}

\date{Accepted 2021 Jan. 16. Received 2020 Dec. 8; in original form 2021 Jan. 16}

\pubyear{2021}

\begin{document}
\label{firstpage}
\pagerange{\pageref{firstpage}--\pageref{lastpage}}
\maketitle

\begin{abstract}
We investigate the impact of asymmetric neutrino-emissions on explosive nucleosynthesis in core-collapse supernovae (CCSNe) of progenitors with a mass range of 9.5 to 25$\Ms$. 
We perform axisymmetric, hydrodynamic simulations of the CCSN explosion with a simplified neutrino-transport, in which anti-correlated dipolar emissions of $\nue$ and $\nueb$ are imposed.
We then evaluate abundances and masses of the CCSN ejecta in a post-processing manner.
We find that the asymmetric $\nu$-emission leads to the abundant ejection of $p$- and $n$-rich matter in the high-$\nue$ and -$\nueb$ hemispheres, respectively.
It substantially affects the abundances of the ejecta for elements heavier than Ni regardless of progenitors, although those elements lighter than Ca are less sensitive. 
Based on these results, we calculate the IMF-averaged abundances of the CCSN ejecta with taking into account the contribution from Type Ia SNe. 
For $m_{\rm asy} = 10/3\%$ and $10\%$, where $m_{\rm asy}$ denotes the asymmetric degree of the dipole components in the neutrino emissions,
the averaged abundances for elements lighter than Y are comparable to those of the solar abundances,
whereas those of elements heavier than Ge are overproduced in the case with $m_{\rm asy} \ge 30\%$.
Our result also suggests that the effect of the asymmetric neutrino emissions is imprinted in the difference of abundance ratio of [Ni/Fe] and [Zn/Fe] between the high-$\nue$ and -$\nueb$ hemispheres, 
indicating that the future spectroscopic X-ray observations of a CCSN remnant will bring evidence of the asymmetric neutrino emissions if exist.
\end{abstract}






\section{Introduction}\label{sec:intro} 

A core-collapse supernova (CCSN) is an energetic explosion of a massive star with a mass of $\ge 8\Ms$, whose explosion energy reaches to $10^{51} \rm erg$ for a typical CCSN.
Although the explosion mechanism is still uncertain, decades of modelling efforts with improving the input physics and numerical techniques  made significant progress in our understanding of CCSN dynamics; 
indeed, we have witnessed that many multi-dimensional(D) simulations by modern CCSN codes in both 2D~\citep{2016ApJ...825....6S, 2018MNRAS.477.3091V, 2018ApJ...854..136N, 2019ApJ...878...13P, 2020ApJ...902..150H, 2020arXiv200914157B} 
and 3D~\citep{2015ApJ...807L..31L, 2016ApJ...831...98R, 2018MNRAS.477L..80K, 2019MNRAS.482..351V, 2019MNRAS.490.4622N, 2019MNRAS.484.3307M, 2019ApJ...873...45G, 2019PhRvD.100f3018W, 2020MNRAS.491.2715B, 2020MNRAS.492.5764N, 2020arXiv201002453P, 2020arXiv201010506B} have demonstrated successful explosions.
Although those results should be still considered provisional, the community has reached the consensus that the CCSN dynamics has strongly influenced on the dimension,
indicating that their observational consequences such as nucleosynthesis in the ejecta need to be also analysed in multi-D, otherwise, the qualitative trend may be missed. 
In this paper, we consider the explosive nucleosynthesis under the multi-D treatments of both matter-dynamics and neutrino transport, albeit approximately.

CCSNe are also known as a cosmic factory of heavy elements synthesized via various channels of hydrostatic nuclear burning and explosive nucleosynthesis during the development of CCSN explosions.
Extensive studies on the nucleosynthesis in massive stars and CCSNe have been performed, based on spherically symmetric models.
The nucleosynthetic computations are carried out on hydrostatic stellar evolution codes from the main sequence to the pre-SN stage, 
and then they are taken over by hydrodynamic simulations for the subsequent CCSN explosion phase, in which a thermal bomb- or a piston-prescription has been usually adapted to trigger the explosion~\citep{1995PThPh..94..663H, 1995ApJS..101..181W, 1996ApJ...460..408T, 2002RvMP...74.1015W, 2002ApJ...576..323R, 2006ApJ...647..483L, 2010ApJ...724..341H, 2018ApJS..237...13L}.
More systematic and elaborated studies in spherical symmetry have been initiated for large numbers of massive stars with a spherical hydrodynamic code
with approximate neutrino transport and a simplified neutrino-core model for the explosion phase~\citep{2012ApJ...757...69U, 2015ApJ...806..275P, 2016ApJ...821...38S}.
Multi-D effects on explosive nucleosynthesis in CCSN have been studied with a 2D hydrodynamic code in a thermal bomb manner~\citep{1997ApJ...486.1026N, 1998ApJ...492L..45N, 2000ApJS..127..141N}
and examined for a more energetic, aspherical SN, or {\it hypernovae}~(see, e.g., \citet{2003ApJ...598.1163M, 2006ApJ...647.1255N}).
{\it r}-process nucleosynthesis in magneto-hydrodynamic (MHD)-driven CCSNe and collapsar-jets, which would be rare events, 
has been investigated with multi-D MHD codes~(see, e.g., \citet{2006ApJ...642..410N, 2007ApJ...656..382F, 2008ApJ...680.1350F, 2009PThPh.122..755O, 2012ApJ...750L..22W, 2015ApJ...810..109N, 2017ApJ...836L..21N, 2018ApJ...864..171M, 2018MNRAS.477.2366H, 2019Natur.569..241S, 2020ApJ...902...66M, 2020arXiv201002227R}).
In the context of neutrino-driven CCSNe, the explosive nucleosynthesis has been also examined for multi-D simulations of CCSNe with approximated neutrino transport~\citep{2005ApJ...623..325P, 2006ApJ...644.1028P, 2011ApJ...738...61F, 2011ApJ...726L..15W, 2013ApJ...774L...6W, 2013ApJ...767L..26W, 2017ApJ...843....2H, 2017ApJ...842...13W, 2018JPhG...45a4001E, 2018ApJ...852...40W, 2020arXiv200812831S}.

Asymmetric neutrino emissions accompanied by lepton-emission self-sustained asymmetry (LESA) have been observed in recent multi-D CCSN simulations with detailed neutrino transport 
among different groups~\citep{2014ApJ...792...96T, 2018ApJ...865...81O, 2019MNRAS.482..351V}
and the emissions with the NS kick are also revealed in a 2D axisymmetric CCSN simulation with full Boltzmann neutrino transport~\citep{2019ApJ...880L..28N}.
Those coherent asymmetric neutrino emissions have a potential influence on the explosive nucleosynthesis in the CCSN ejecta; hence, we studied them for an SN1987A-like progenitor with a mass of 19.4$\Ms$ in \citet{2019MNRAS.488L.114F}.
Employed with an axisymmetric hydrodynamic code with a simplified neutrino transport and a neutrino-core model~\citep[see also][]{2012ApJ...757...69U, 2016ApJ...821...38S}, 
we have shown that the asymmetric emissions of $\nu$ tend to yield larger amounts of $n$- and $p$-rich ejecta in the hemisphere of the higher $\nueb$ and $\nue$ emissions, respectively.
For small magnitudes of the neutrino asymmetry, or $10/3\%$ and $10\%$,
abundances of elements heavier than Zn are comparable to or slightly larger than those of the solar abundances.
On the other hand, for larger asymmetric cases ($\ge 30\%$), 
the ejecta have too many elements compared to the solar abundances due to the larger amounts of the $n$-rich ejecta.

In our previous study, however, we focused on the case with a single progenitor, which is not enough to make a detailed comparison to the solar abundances. 
To this end, we need to cover a wide range of progenitors on our nucleosynthetic computations, with which we calculate IMF-averaged abundances. 
It should be also mentioned that the comprehensive study of the progenitor dependence is inevitable to compare our results with future observations of SN remnants (SNRs).
In this paper, we perform 2D hydrodynamic simulations of the CCSN explosion and nucleosynthetic calculations for six progenitors with a mass range from 9.5 to 25$\Ms$, in addition to the 19.4$\Ms$ progenitor of our previous work. 
Note that we change the asymmetric degree of neutrino emissions in each progenitor, thus we have 35 models in total.
We confirm that the overall trend is qualitatively in line with that found in our previous paper, although there are some diversities among progenitors, which we delve into in this study.

This paper is organized as follows. In section 2, we describe the methods and models for numerical simulations of 2D axisymmetric simulations of CCSN explosions and nucleosynthetic calculations.
In section 3, we first summarize the essential trend found in our previous study for the $19.4\Ms$ progenitor
and then present the results of 2D CCSN simulations and nucleosynthetic computations for the other progenitors.
We then discuss asymmetric distributions of the ejecta with particularly focusing on compositional differences in section 4, which provides an important insight towards spectroscopic X-ray observations on SNRs.
Finally, we will summarize our conclusion in section 5.

\section{Method \& model}\label{sec:method} 

\subsection{Axisymmetric simulations of CCSN explosion}\label{sec:SN explosion} 

As in our previous study~\citep{2019MNRAS.488L.114F},
we perform hydrodynamic simulations from core collapse to SN explosion of massive stars, employed with two codes, the GR1D code~\citep{2015ApJS..219...24O} 
and a modified Zeus 2D code~\citep{1992ApJS...80..753S,1992ApJS...80..791S, 2006ApJ...641.1018O, 2007ApJ...667..375O, 2011ApJ...738...61F},
in which a simplified $\nu$ transport, or a light-bulb scheme, is adopted and appropriate weak interactions are taken into account~(see detailed in Appendix of \citet{2011ApJ...738...61F}).
We follow matter evolution from the core collapse to the shock stole with the GR1D code in spherical symmetry, and then the results are mapped in the polar coordinate $(r, \theta)$ on the Zeus 2D code.
The computational domain in 2D simulations is from $50\rm$ to $50,000\km$ in $r$ and $0 \le \theta \le \pi$, covered with $(500, 128)$ meshes.
The central part ($\le 50\km$) of the proto-neutron star (proto-NS) is excised and is treated as the central point source with a mass of $M$,
which continuously increases due to mass accretion through the inner boundary ($50\km$) of the computational domain of our 2D simulations. 
At the time of remapping from 1D to 2D, the mass of a central point source is $1.2\Ms$, 
and we add non-radial ($l=1$) perturbations on radial velocities, whose magnitudes are $1\%$, to break the spherical symmetry.

In the present study, neutrinos are assumed to be emitted from the neutrinospheres with a thermal spectrum.
We assume that the neutrino temperatures are spherical symmetric but luminosities have dipole components;
\begin{eqnarray}
L_{\nue} &=& L_{\nue, \rm ave} (1 +m_{\rm asy} \cos \theta), \label{eq:asymmetric ne luminosity} \\
L_{\nueb} &=& L_{\nueb, \rm ave} (1 -m_{\rm asy} \cos \theta), \label{eq:asymmetric neb luminosity}
\end{eqnarray}
where $L_{\nue}$ and $L_{\nueb}$ are luminosity of $\nue$ and $\nueb$ and $L_{\nue, \rm ave}$ and $L_{\nue, \rm ave}$ are angular-averaged $L_{\nue}$ and $L_{\nueb}$, respectively.
Here the evolution of $L_{\nue, \rm ave}$, $L_{\nueb, \rm ave}$ and $\nu$ temperatures are evaluated with a $\nu$-core model from the mass accretion rate at the inner boundary of the computational domain, 
as in \citet{2012ApJ...757...69U, 2016ApJ...821...38S} but with some modifications (Appendix \ref{sec: neutrino core model}).
As in our previous work~\citep{2019MNRAS.488L.114F}, we have tuned two parameters of the $\nu$-core model 
so that the $19.4\Ms$ progenitor explodes as SN1987A-like (the explosion energy, $E_{\rm exp}$, $\sim 10^{51} \rm erg$ 
and the ejected mass of \nuc{Ni}{56}, $M$(\nuc{Ni}{56}), of $(0.07-0.08) \Ms$; ) for cases of spherical $\nu$ emission ($m_{\rm asy} = 0\%$).
We run the simulations by fixing those tunned parameters but changing $m_{\rm asy}$ as 0\%, 10/3\%, 10\%, 30\%, and 50\%.
We note that the typical asymmetry of neutrino emissions may be $m_{\rm asy} \le 10 \%$~\citep[see e.g., ][]{2014ApJ...792...96T, 2019ApJ...880L..28N, 2019MNRAS.489.2227V}, although it is not definitive;
hence, we investigate cases with higher asymmetric, neutrino-emissions in this study.

In the present study, we investigate the impact of neutrino asymmetry on explosive nucleosynthesis in CCSNe of progenitors~\citep{2015ApJ...810...34W, 2002RvMP...74.1015W} 
with a mass of $9.5\Ms, 11.2\Ms, 13.0\Ms, 15.0\Ms, 17.0\Ms$, and $25.0\Ms$ in addition to the progenitor of $19.4\Ms$,
whose detailed nucleosynthetic results are presented in our previous study~\citep{2019MNRAS.488L.114F} (see also \S 3.1).
 \begin{figure}
  \begin{center}
   \includegraphics[scale=0.6]{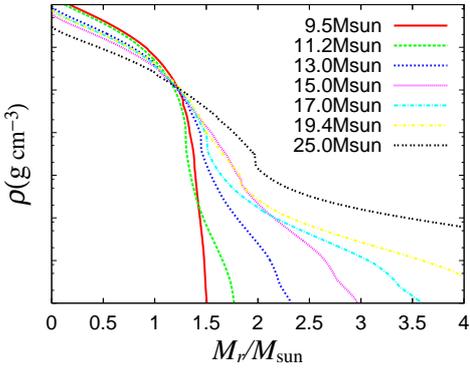}
  \end{center}
  \caption{
  Mass-density profiles as a function of a mass-coordinate $M_r$ for the progenitors adopted in this study. }
  \label{fig:rho-Mr}
 \end{figure}
Figure \ref{fig:rho-Mr} shows density profiles of the progenitors of $\Mms = 9.5\Ms, 11.2\Ms, 13.0\Ms, 15.0\Ms, 17.0\Ms, 19.4\Ms$, and $25.0\Ms$ as a function of a mass-coordinate $M_r$.
The 9.5(15.0) $\Ms$ progenitor has a density profile very similar to that of the 11.2(19.4)$\Ms$ progenitor, 
but shallower at the mass coordinate, $M_r$, $\ge 1.5\Ms$ ($2.0\Ms$).
At $M_r \ge 1.5\Ms$, 
the density of the 13.0 and 17.0$\Ms$ progenitors lie between those of the light (9.5 and 11.2$\Ms$) progenitors and those of the 15.0 and 19.4$\Ms$ progenitors,
while the density is much higher for the 25.0$\Ms$ progenitor.
As we shall discuss below, the accretion component of neutrino luminosity is directly associated with the density profile at the onset of collapse,
indicating that the difference of the density profile is responsible for the progenitor dependence of the impact of asymmetric neutrino emissions on the explosive nucleosynthesis (see Sec.~\ref{sec:results} for more details).

It should be mentioned that a mass cut-off due to a black hole formation needs to be assumed in order to compute the IMF-averaged abundances.
We refer \citet{2013ApJ...769...99B,2018ApJ...852..101S} which suggest that the mass cut-off with $\ge 25-40\Ms$ is appropriate.
A maximum mass of $25\Ms$, therefore, seems to be adequate for the current study.
It should be mentioned that the IMF-averaged [X/Fe] is insensitive to the mass cut-off in spherically symmetric neutrino emission models as long as it is $\ge 25-40\Ms$ (see Fujimoto et al. (2021) in preparation);
hence our choice of the mass cut-off would give fewer impacts on our present results.
We also note that the seven progenitors employed in this study are enough to compute the averaged IMF abundances,
since the relative contribution of the averaged [X/Fe] over the seven progenitors are comparable to that over more than 20 progenitors from $9.5\Ms$ to $25.0\Ms$
in our spherically symmetric neutrino emission models (Fujimoto et al. (2021) in preparation).

\begin{figure*}
  \begin{center}
   \includegraphics[scale=0.7]{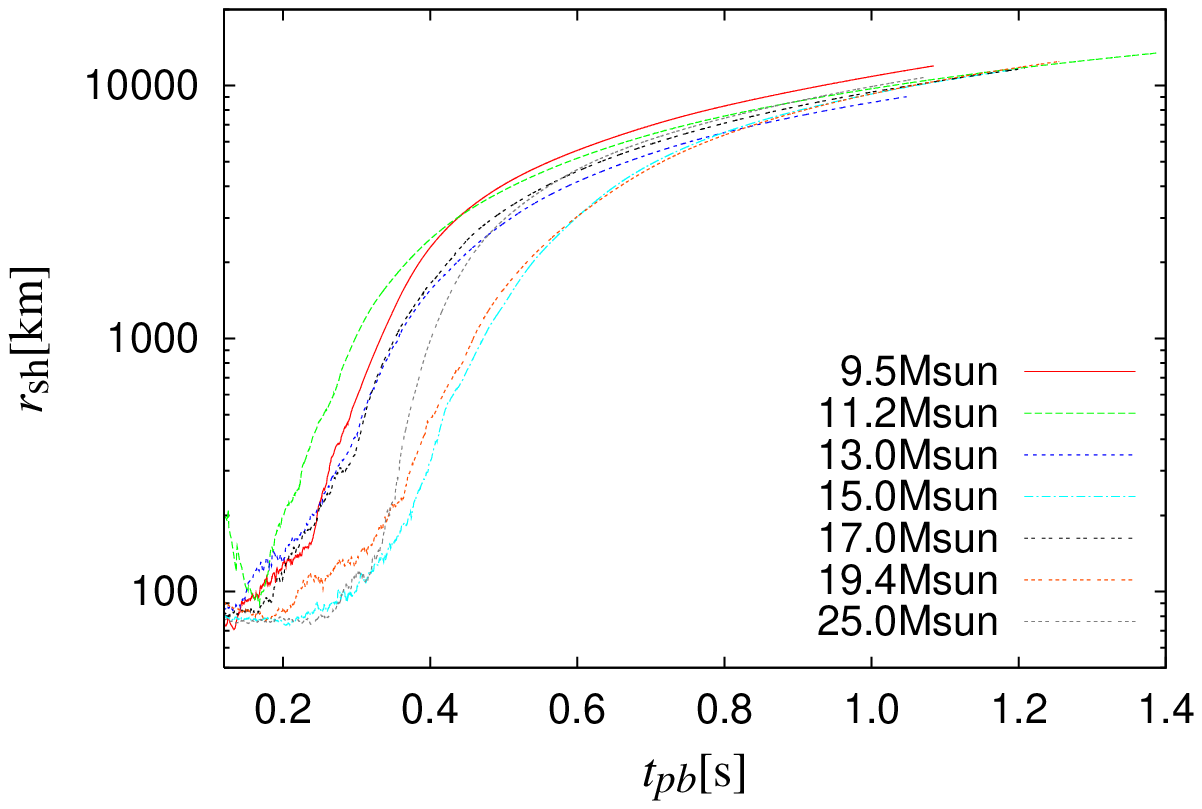}
   \includegraphics[scale=0.65]{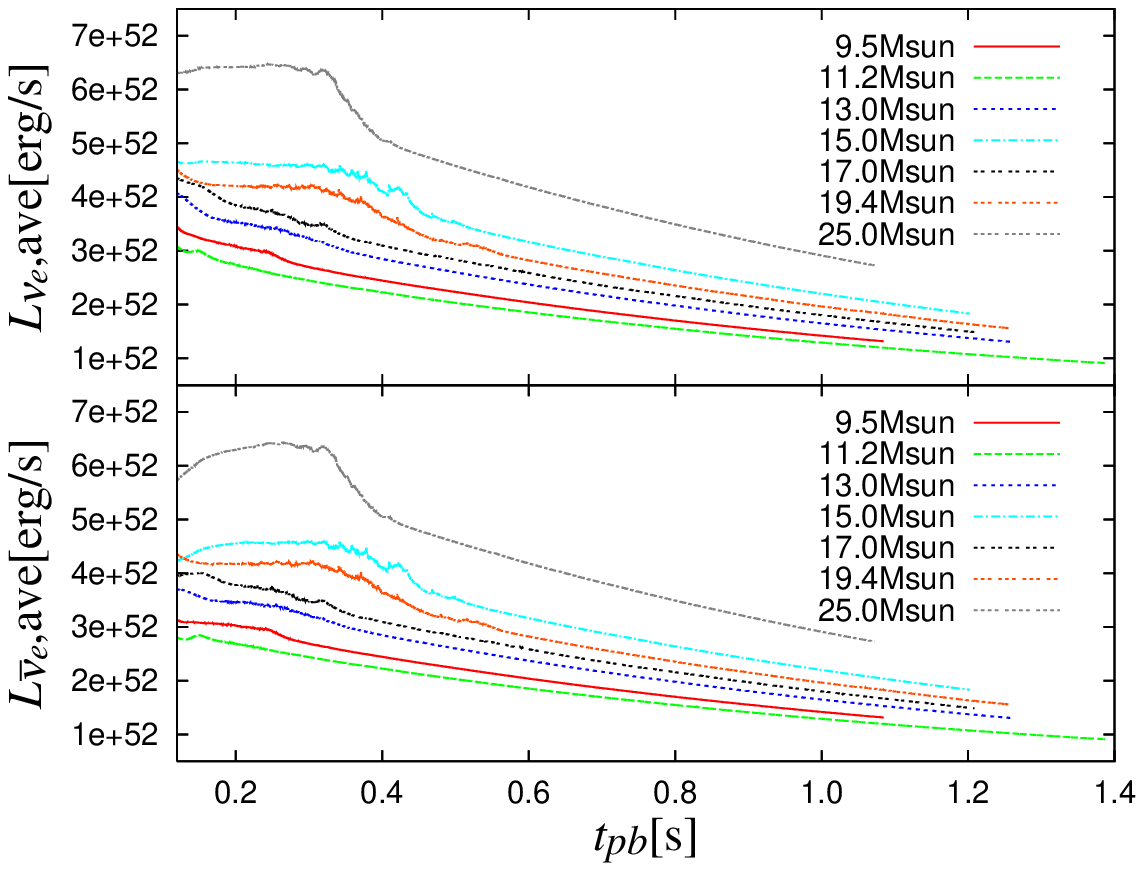}
  \end{center}
  \caption{
  Time evolution of the angular-averaged shock radii $r_{\rm sh}$ (left panel), and $L_{\nue,\rm ave}$ (top right panel), and $L_{\nueb,\rm ave}$ (bottom right panel) as a function of $t_{\rm pb}$.
  We note that they are less sensitive to $m_{\rm asy}$; hence, we only display the results with $m_{\rm asy}=0\%$.
  }
  \label{fig:Rsh-Lnu}
 \end{figure*}

We terminate our CCSN simulations when the shock wave reaches to $\sim 10,000\km$.
Figure \ref{fig:Rsh-Lnu} shows the time evolution of angular-averaged shock radii, $r_{\rm sh}$ (left panel), $L_{\nue,\rm ave}$ (top right panel), and $L_{\nueb,\rm ave}$ (bottom right panel)
for case with symmetric neutrino emission ($m_{\rm asy}=0\%$) and for progenitors of $9.5\Ms, 11.2\Ms, 13.0\Ms, 15.0\Ms, 17.0\Ms, 19.4\Ms$, and $25.0\Ms$
as a function of the time from the core bounce ($t_{\rm pb}$).
Note that the evolution of $r_{\rm sh}$, neutrino luminosities, and temperatures weakly depend on $m_{\rm asy}$; hence, we display only the results with $m_{\rm asy}=0$ in these panels.
For the light progenitors: $9.5\Ms$, $11.2\Ms$, $13.0\Ms$ and, $17.0\Ms$,
shock revivals occur at relatively early time ($\sim$ 200\ms),
while the shock wave for the other progenitors ($15.0\Ms$, $19.4\Ms$, and $25.0\Ms$) revives at a later phase $\sim$ 300\ms.
Neutrino luminosities are low for the light progenitors of $9.5\Ms$ and $11.2\Ms$, whereas they are higher for heavier progenitors due to larger mass accretion rates.
In particular, for progenitors with high compactness ($\Mms = 15.0\Ms$, $19.4\Ms$, and $25.0\Ms$),
the luminosities and temperatures are remarkably higher than other models.

\subsection{Nucleosynthetic calculation}\label{sec:nucleosynthesis}

We follow Lagrangian thermodynamic histories of SN ejecta for nucleosynthetic computation with a tracer particle method~\citep{1997ApJ...486.1026N, 2010MNRAS.407.2297S, 2017ApJ...843....2H},
in which tracer particles evolve in accordance with the fluid-velocity of the CCSN simulations while storing their physical quantities such as density, temperature, and electron fraction, $Y_e$.
We distribute 6,000 tracer particles in the regions from $1,000$km to $10,000$km (or an O-rich layer) at the beginning of the simulations, 
with being adaptively weighted to the mass of a layer where the particle is located.
By virtue of the adaptive mass of tracer particles, the highest resolutions in the particle mass become $\sim (10^{-5}$-$10^{-4})\Ms$ in the present study.
We made a sensitivity study of nuclear abundances to the particle numbers in \citep{2011ApJ...738...61F}, and a reasonable convergence is achieved at 6,000 particles; indeed,
we found that the differences in ejected masses of nuclei between 3,000 and 6,000 particles cases are less than $\sim$ 1\%, hence, we adopted 6,000 particles in this study.
In our models, $\sim$ 3,000-4,500 particles are ejected by SN explosions, 
whose masses weakly depend on $m_{\rm asy}$ and are evaluated to be
0.09$\Ms$, (0.20-0.22)$\Ms$, (0.35-0.36)$\Ms$, (0.44-0.47)$\Ms$, (0.53-0.56)$\Ms$, (0.52-0.58)$\Ms$, and (0.78-0.81)$\Ms$, 
for progenitors with $(9.5, 11.2, 13.0, 15.0, 17.0, 19.4$, and $25.0)\Ms$.

We follow the abundance evolution of 2,488 nuclides from $n$, $p$ to \nuc{Nd}{196} in the CCSN ejecta by employing the Lagrangian thermodynamic histories of the ejecta 
and a nuclear reaction network as in our previous study~\citep{2019MNRAS.488L.114F}.
Note that, when the ejecta become hotter than $9 \times 10^9 \K$, we set the abundances of the nuclides to be the chemical composition in nuclear statistical equilibrium. 
We also take into account abundance change via neutrino interactions for He and nuclei from C to Kr as in \citet{2011ApJ...738...61F}.
In the nucleosynthetic computations, we recompute $Y_e$ following the weak interactions employed in the nuclear reaction network, albeit possessing $Y_e$ data from the hydrodynamic simulations.
This is mainly because more elaborate weak reactions are incorporated in the nuclear reaction network than those used in 2D hydrodynamic simulations~\citep{2011ApJ...738...61F}, 
and the recomputation is necessary in particular for reversed-$L_\nu$ cases (see below) in nucleosynthetic computations.

We note that the deviation of the neutrino spectra from a Fermi-Dirac distribution has a negligible effect on the composition of the SN ejecta via $\nu$-processes.
The impact was well investigated by the two simulations in \citet{2019ApJ...876..151S};
a simulation of a piston-driven CCSN explosion of a $27\Ms$ progenitor and $E_{\rm exp} = 1.2 \times 10^{51} \erg$ with a quasi-thermal energy distribution of $\nu$
and a spherically symmetric, artificially exploded CCSN simulation of the $27\Ms$ progenitor, whose $\nu$ luminosities and spectra evaluated with a hydrodynamic code,
in which an approximated neutrino transport is taken into account.
We, hence, ignore the effect of deviations from the thermal spectrum in neutrino spectra in this study.

We find that in the region, $r_{\rm cc} \le 10,000\km$, the peak temperature of the SN shock wave becomes higher than $\sim 1 \times 10^9 \K$, where nuclei heavier than C are abundantly synthesized.
Here $r_{\rm cc}$ is the radius of the ejecta at the onset of the core collapse.
We therefore conduct nucleosynthetic computation only for the tracer particles, which are located $r_{\rm cc} \le 10,000\km$.
For the initial abundances of the particles, 
we take abundances of 20 nuclei in the progenitors at the onset of the core collapse. The abundances are taken from the results of stellar evolution in~\citet{2002RvMP...74.1015W}.
For $15.0\Ms$ and $25.0\Ms$ progenitors, we also adopt alternative abundance data of more than 1,300 nuclei~\citep{2002ApJ...576..323R},
which consist of those with $Z \le 83$ including odd $Z$ nuclei as well as $s$-process nuclei produced via the weak $s$-process.
The chemical composition of the outer ejecta with $r_{\rm cc} > 10,000\km$ is set to be the abundances in the progenitors at the onset of the core collapse,
ignoring the $\gamma$-processes in $p$-process layers~\citep{1995A&A...298..517R} and the $\nu$-processes in the outer layers~\citep{2019ApJ...876..151S}.
We note that the ejected masses from the region $r > r_{\rm cc}$ are $7.73\Ms$, $9.30\Ms$, $9.49\Ms$, $10.5\Ms$, $11.7\Ms$, $13.0\Ms$, and $9.75\Ms$
for progenitors with $9.5\Ms, 11.2\Ms, 13.0\Ms, 15.0\Ms, 17.0\Ms$, $19.4\Ms$, and $25.0\Ms$.

Recent multi-D CCSN simulations reveal the correlation between the directions of stronger shock and the hemisphere with high-$\nueb$ emissions (the high-$\nueb$ hemisphere)
(see, e.g., \citet{2014ApJ...792...96T, 2019ApJ...880L..28N, 2019MNRAS.489.2227V}).
In some of our models, however, the direction of the high-$\nueb$ hemisphere does not correlate with that of the stronger shock expansion. 
It is attributed to the fact that the shock revival takes place in stochastic directions, meanwhile we set a priori the dipole direction of neutrino asymmetry.
We find that ten models are mis-correlated; ($M_{\rm ms}/\Ms$, $m_{\rm asy}/\%$) = $(9.5, 10)$, $(9.5, 30)$, $(13.0, 10)$, $(13.0, 30)$, $(13.0, 50)$, $(15.0, 30)$, $(17.0, 10)$, $(17.0, 30)$, $(17.0, 50)$, and $(25.0, 10)$.
For the nucleosynthetic calculation in these mis-correlated models, 
we retain the matter data but reverse the direction of $\nu$ asymmetry in nucleosynthetic computations.
We discuss the impact of the reversed-$L_\nu$ procedure in Appendix \ref{sec:reversed-Lnu}, and confirm that it is a reasonable prescription to obtain qualitatively accurate results.

\section{results}\label{sec:results}

\subsection{Summary of our results in previous paper}\label{sec:previos work}

We briefly summarize the essential results of our previous study~\citep{2019MNRAS.488L.114F}: the impact of asymmetric neutrino emissions on explosive nucleosynthesis for a CCSN of the $19.4\Ms$ progenitor.
As shown in Sec. \ref{sec:SN explosion}, 
the progenitor is tuned to reveal the SN1987A-like explosion with $E_{\rm exp} \sim 10^{51} \rm erg$ and $M($\nuc{Ni}{56}$)\sim (0.07-0.08) \Ms$.
In our model, the shock wave is revived at $\sim 0.3$ s after the core bounce (Fig. \ref{fig:Rsh-Lnu}) and then it expands quasi-spherically.
The explosion exhibits the aspherical distribution of the density, temperature, and entropy of neutrino-driven inner ejecta.
Although the fluid-dynamics is less sensitive to $m_{\rm asy}$, the distribution of $Y_e$ of SN ejecta strongly depends on $m_{\rm asy}$ due to a different amount of $\nu$ absorption 
through the different degree of the dipolar and anti-correlated $\nu$ emissions (Eqs. (1) and (2));
In the hemisphere of $\pi/2 < \theta \le \pi$ (the high-$\nueb$ hemisphere), the asymmetric emissions of $\nu$ tend to yield larger amounts of $n$-rich ejecta ($Y_e < 0.49$),
while $p$-rich matter ($Y_e > 0.51$) are ejected in the opposite hemisphere of the higher $\nue$ emissions ($0 \le \theta < \pi/2$; the high-$\nue$ hemisphere).
See also Fig. 1 in \citet{2019MNRAS.488L.114F} for more details.

For larger asymmetric cases with $m_{\rm asy}\ge 30\%$, the larger amounts of the $n$-rich ejecta are produced in the high-$\nueb$ hemisphere.
As a result, there are too many elements heavier than Zn compared to the solar abundances~(Fig. 4 in \citet{2019MNRAS.488L.114F}).
For cases with small $m_{\rm asy}$ ($10/3\%$ and $10\%$), abundances of the elements are comparable to or slightly larger than those of the solar abundances.
We also observed the characteristic features in elemental distribution;
abundances lighter than Ca are insensitive to $m_{\rm asy}$ and the production of Ni, Zn, and Ge is much larger in the $n$-rich ejecta in the high-$\nueb$ hemisphere 
than those in the $p$-rich ejecta in the high-$\nue$ hemisphere even for smaller asymmetric cases with $\le 10\%$.
With these results in mind, let us delve into the progenitor dependence, which is the main subject in this paper.

\subsection{Abundances of SN ejecta}\label{sec:abundances}

\begin{figure*}
 \includegraphics[scale=1.0]{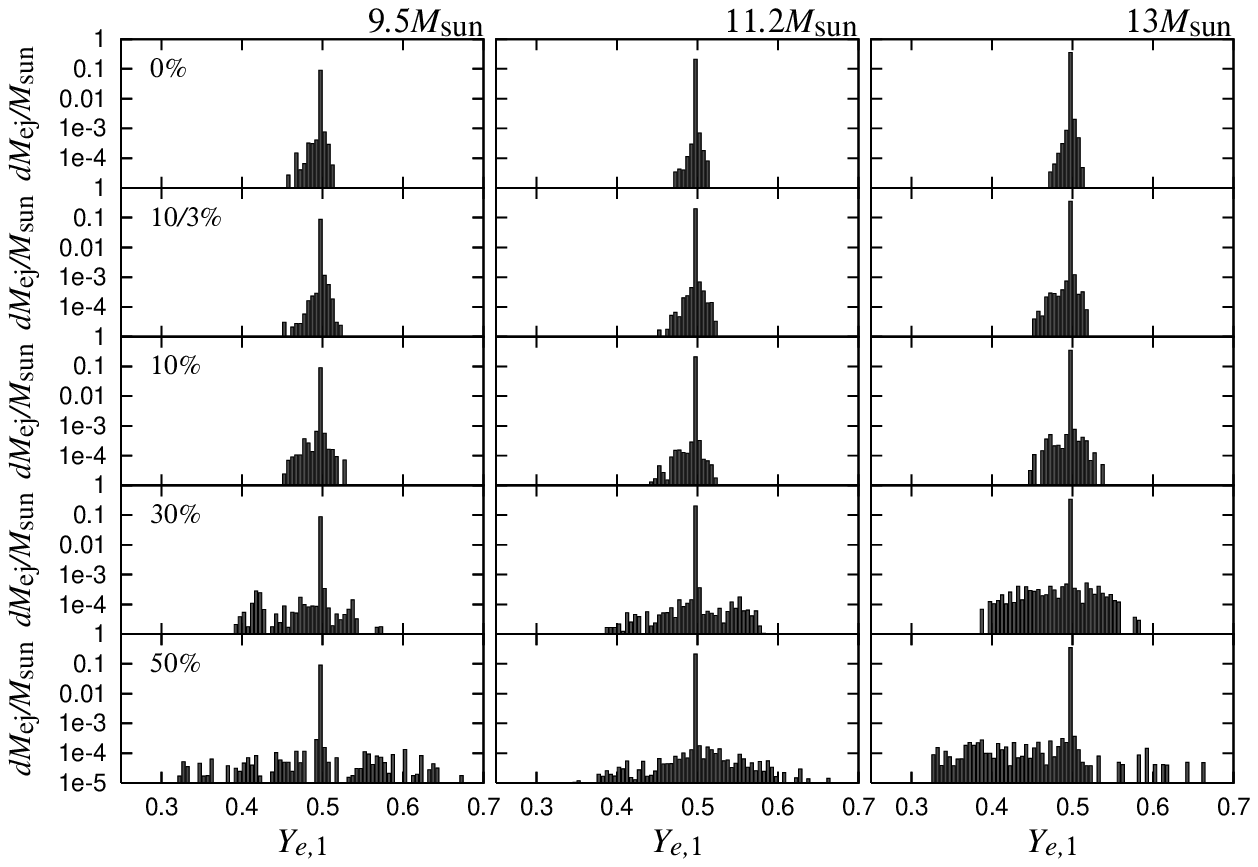}
 \includegraphics[scale=1.0]{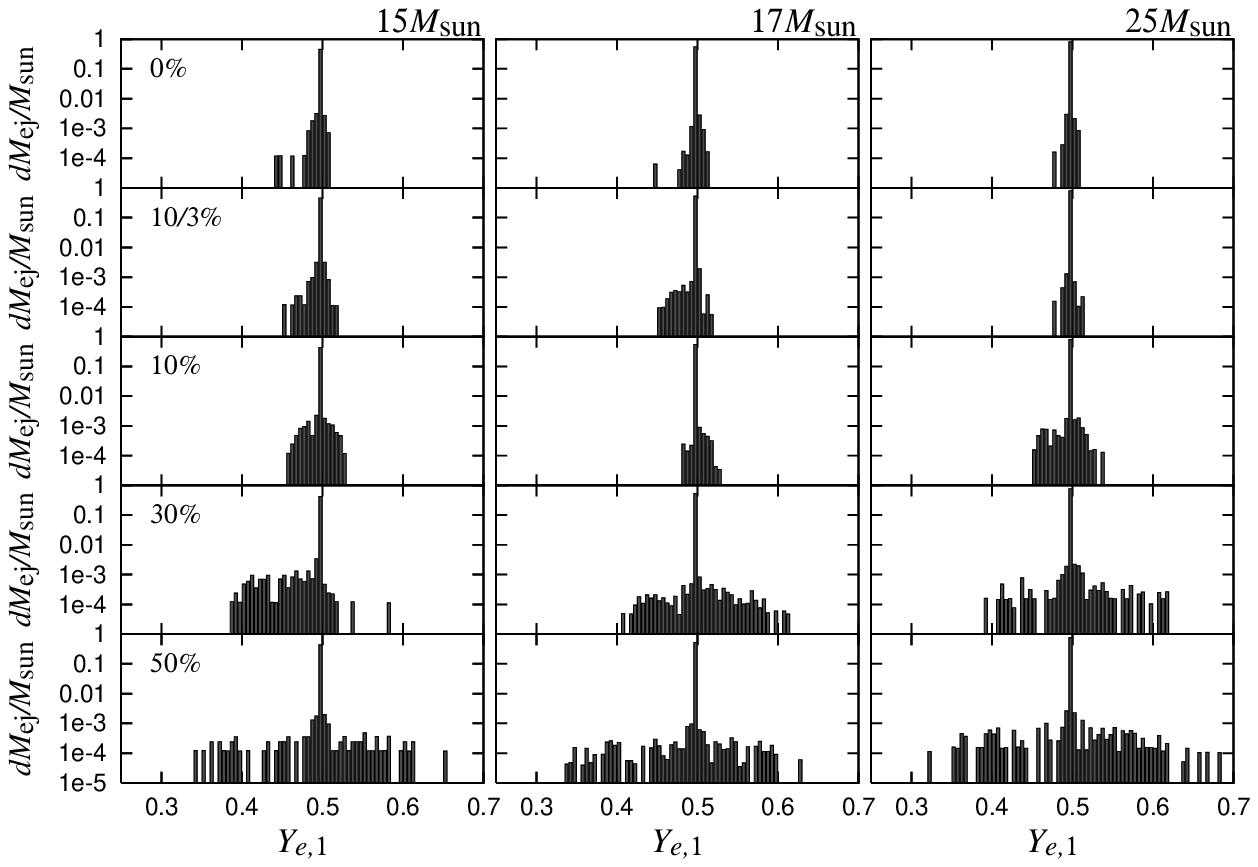}
 \caption{
 Progenitor dependence of mass profiles of $dM_{\rm ej}$ in $Y_{\rm e,1}$ of the ejecta from the inner region ($r_{\rm cc} \le 10,000\km$). 
 We display the results of all progenitors except for $\Mms = 19.4\Ms$ \citep[see Fig. 3 in][for $\Mms = 19.4\Ms$ model]{2019MNRAS.488L.114F}.
 The asymmetric degree of the neutrino emissions, $m_{\rm asy}$, is increased from top to bottom in each panel.
 }
 \label{fig:dist-ye-Mms}
\end{figure*}

Figure \ref{fig:dist-ye-Mms} shows mass profiles of $dM_{\rm ej}$ in $Y_{\rm e,1}$ of the ejecta from the inner region ($r_{\rm cc} \le 10,000\km$)
for the progenitors with $\Mms = 9.5\Ms, 11.2\Ms, 13.0\Ms, 15.0\Ms, 17.0\Ms$, and $25.0\Ms$, 
and for cases with $m_{\rm asy}=$ 0\%, 10/3\%, 10\%, 30\%, and 50\%. 
Here $Y_{\rm e,1}$ is the electron fraction evaluated when the temperature is equal to $1 \times 10^9 \rm K$ during the ejection
and $dM_{\rm ej}$ is a mass of ejecta integrated with a bin of $dY_{\rm e,1} = 0.005$.
We note that $Y_e$ of the ejecta chiefly changes through the neutrino absorption near the proto-NS during the ejection,
implying that it freezes out at the place with $Y_e \sim Y_{\rm e,1}$  where it is far away from proto-NS in this phase~\citep[see, e.g., Fig.4 in][]{2011ApJ...738...61F}.
The ejecta goes through the adiabatic cooling with expansions but $Y_e$ re-increases from a certain point due to subsequent $\beta$-decays of unstable nuclei in the ejecta.
For larger $m_{\rm asy}$, 
$Y_{\rm e,1}$ distributions are widely varying, indicating that mass fractions of $n$- and $p$-rich ejecta become larger 
due to larger $\nueb$ and $\nue$ absorptions on nucleon in the ejecta.
We also find that masses of $n$-rich ejecta tend to be larger with larger amounts of the inner ejecta and with larger $\nueb$ luminosities.
It should be mentioned, however, that the mass of $n$-rich ejecta is not a monotonic function of the progenitor mass; indeed, they are smaller in $17.0\Ms$ ($11.2\Ms$) than in $15.0\Ms$ ($9.5\Ms$).
It is attributed to the fact that the mass density profile of progenitors just prior to the onset of gravitational collapse is a non-monotonic order to the zero age main-sequence mass. 
This can be seen in Fig.~\ref{fig:rho-Mr}; there are regions where the density becomes lower in $17.0\Ms$ ($11.2\Ms$) than $15.0\Ms$ ($9.5\Ms$),
which results in decreasing accretion components of neutrino luminosity (see also Fig.\ref{fig:Rsh-Lnu}).
As a result, the impact of asymmetric neutrino emissions becomes weak, implying that the production of $n$-rich ejecta is less efficient.

\begin{figure*}
 \includegraphics[scale=1.25]{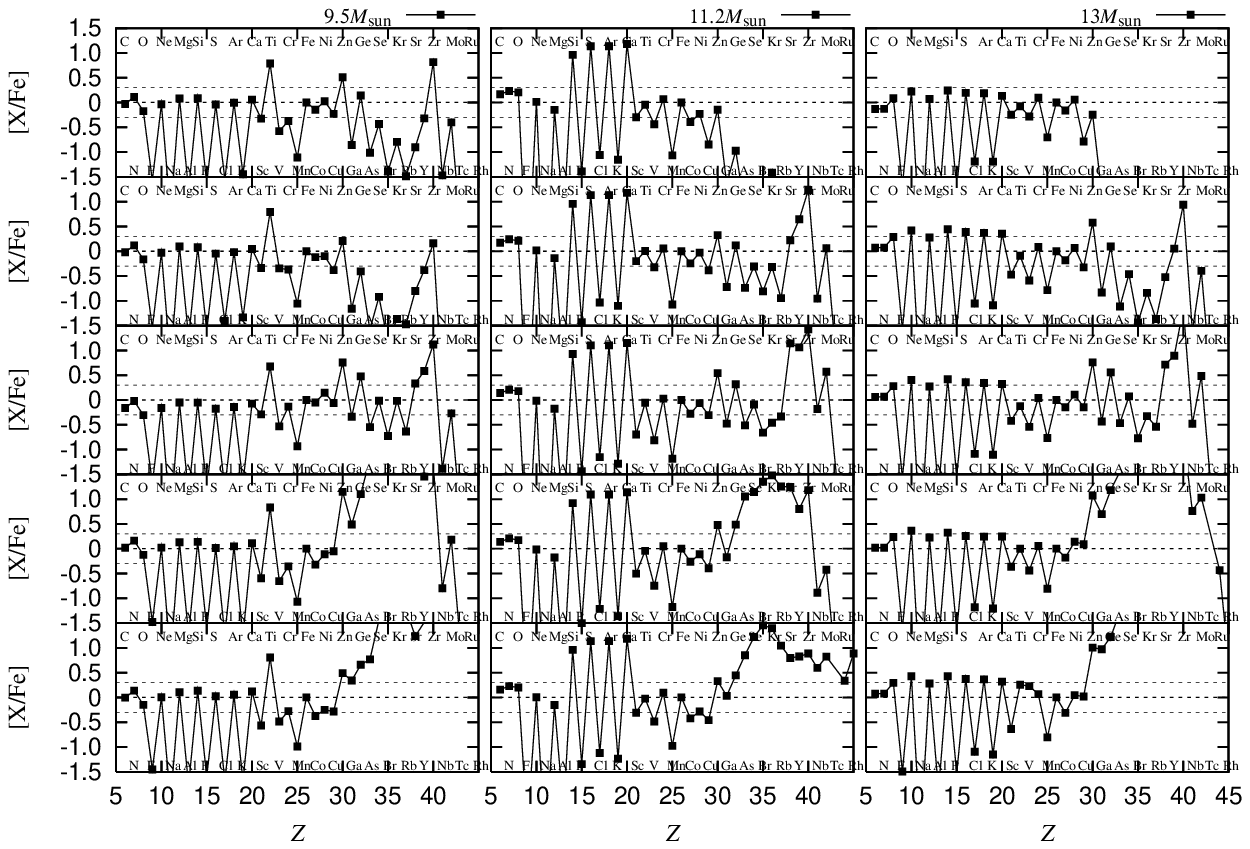}
 \includegraphics[scale=1.25]{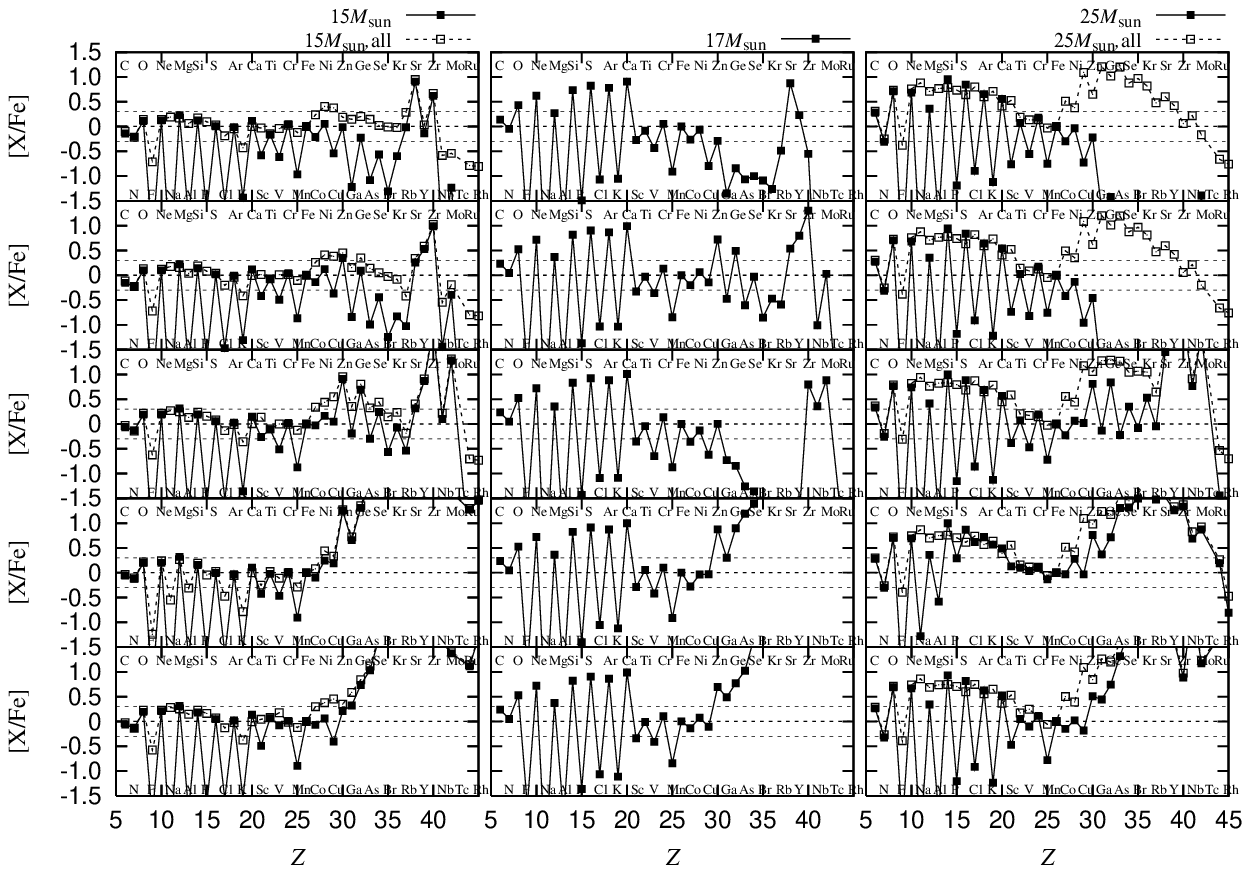}
 \vspace*{-10pt}
 \caption{
 Same as Fig. \ref{fig:dist-ye-Mms} but for [X/Fe]. 
 $m_{\rm asy}$ is 0\%, 10/3\%, 10\%, 30\%, and 50\% from top to bottom in each panel.
 Solid-lines with filled-squares and dashed-lines with open-squares show [X/Fe] for cases with the initial abundances of 20 nuclei 
 in a stellar evolution code~\citep{2002RvMP...74.1015W} and of more than 1,300 nuclei~\citep{2002ApJ...576..323R} (only for the $15.0\Ms$ and $25.0\Ms$ progenitors), respectively.
 We note that the pronounced odd-even patterns of [X/Fe] for elements lighter than Fe would be a numerical artifact due to the small number of the initial abundances. See the text for more details.
 }
 \label{fig:xfe-Mms}
\end{figure*}

Figure \ref{fig:xfe-Mms} shows the composition of the ejecta for our models 
in terms of [X/Fe]~\footnote{[A/B] = $\log \left[ (X_{\rm A}/X_{\rm A, \odot})/(X_{\rm B}/X_{\rm B, \odot})\right]$,
where $X_{\rm i}$ and $X_{\rm i,\odot}$ are a mass fraction of an element $\rm i$ and its solar value~\citep{1989GeCoA..53..197A}, respectively.} 
of the ejecta as a function of an atomic number $Z$ 
for the progenitors with $\Mms = 9.5\Ms, 11.2\Ms, 13.0\Ms, 15.0\Ms, 17.0\Ms$, and $25.0\Ms$
and for cases with $m_{\rm asy}=$ 0\%, 10/3\%, 10\%, 30\%, and 50\%.
We present [X/Fe] for cases with initial abundances of 20 nuclei in the stellar evolution code~\citep{2002RvMP...74.1015W} (solid-lines with filled-squares)
and those for cases with the alternative initial abundances of more than 1,300 nuclei~\citep{2002ApJ...576..323R} (dashed-lines with open-squares) 
only for the $15.0\Ms$ and $25.0\Ms$ progenitors.
We find that [X/Fe] of odd-$Z$ elements and elements with $Z \ge 27$ are largely underproduced if we adopt the initial abundances of the 20 nuclei,
while [X/Fe] of even-$Z$ elements with $Z \le 26$ are comparable.
Hence, the pronounced odd-even patterns of [X/Fe] for elements lighter than Fe would be due to the small number of initial abundances (20 nuclei) in our simulations.
The comparison suggests that the odd-even patterns of [X/Fe] found in other progenitor models may disappear if we employ initial abundances computed with larger nuclear reaction network.
Keeping in mind the systematic error, let us study the progenitor dependence of the impact of asymmetric neutrino emissions on the abundance pattern in the ejecta.

As in case with the $19.4\Ms$ progenitor \citep[see Fig.4 in][]{2019MNRAS.488L.114F}, [X/Fe] for elements lighter than Ca are insensitive to $m_{\rm asy}$.
For symmetric $L_\nu$ cases ($m_{\rm asy} = 0$), elements heavier than Cu ($Z = 29$) other than Sr, Y, and Zr are underproduced except for the progenitor of $9.5\Ms$.
The exceptional trend found in $9.5\Ms$ model is mainly due to the fact that 
the fractions of $n$-rich ejecta are large even in the case with $m_{\rm asy} = 0$ (see top left panel in Fig.\ref{fig:dist-ye-Mms}).
It should be mentioned that isotopes of Sr, Y, and Zr with the neutron number $N=50$ are abundantly produced in $n$-rich ejecta ($Y_{\rm e,1} \le 0.49$).
which appear appreciably in 9.5, 15.0, and 17.0$\Ms$ models even in the case with $m_{\rm asy} = 0$.
For larger $m_{\rm asy}$, on the other hand, [X/Fe] for elements heavier than Zn ($Z = 30$) become larger due to larger fractions of $n$-rich elements.
In particular, for cases with $m_{\rm asy} \ge 30\%$, these elements are overproduced compared with the solar abundances regardless of progenitors.
We also note that Zn is synthesized in the {\it p}-rich ejecta but the contribution to Zn in the ejecta is minor because of the small abundances (mass fraction $< 0.01$; see e.g., Fig.10 in \citet{2018ApJ...852...40W}).

It is also interesting to be mentioned that [Zn/Fe] for the case with $m_{\rm asy} = 10/3\%$ in 9.5$\Ms$ model is less than those for $m_{\rm asy} = 0$, 
indicating that the abundance ratio is not always monotonic function with $m_{\rm asy}$.
This peculiar trend can be understood as follows.
Let us first point out that Zn is chiefly composed by \nuc{Zn}{64} and \nuc{Zn}{66} and abundantly synthesized in the ejecta with $Y_{e,1} \sim 0.45-0.48$ and $\le 0.43$ 
(see, e.g., Fig.10 in \citet{2018ApJ...852...40W}),
i.e., there is a hole between the two $Y_{e,1}$ regions. For the 9.5$\Ms$ model, Zn is mainly produced in the region of $Y_{e,1} \sim 0.45-0.48$ in the case with $m_{\rm asy} = 0$.
With increasing $m_{\rm asy}$ to $10/3\%$, the ejecta tends to be shifted to the region of the lower $Y_{e,1}$ due to the higher $\nueb$ absorptions, 
and some of them get into the hole (see in the top, left figure on Fig.~\ref{fig:dist-ye-Mms}), results in decreasing [Zn/Fe].

\begin{figure}
 \includegraphics[scale=1.1]{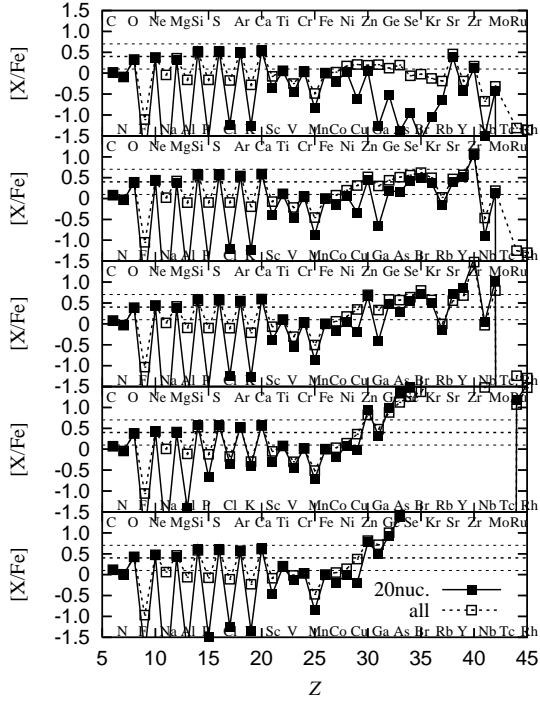}
 \vspace*{-10pt}
 \caption{
 IMF-averaged [X/Fe] of total ejecta for cases with $m_{\rm asy}=$ 0\%, 10/3\%, 10\%, 30\%, and 50\% in panels from top to bottom.
 Solid-lines with filled-squares and dashed-lines with open-squares  represent the cases with the initial abundances 
 of 20 nuclei in a stellar evolution code~\citep{2002RvMP...74.1015W} and of more than 1,300 nuclei~\citep{2002ApJ...576..323R}
 only for the $15.0\Ms$ and $25.0\Ms$ progenitors, respectively.
 }
 \label{fig:xfe-imf}
\end{figure}

We now turn our attention to the comparison between IMF-averaged abundances of our results and the solar one.
We average abundances of the ejecta with the Salpeter IMF weighted by their masses over the seven progenitors of 
$9.5\Ms, 11.2\Ms, 13.0\Ms, 15.0\Ms, 17.0\Ms, 19.4\Ms$, and $25.0\Ms$, setting the minimum and maximum masses of CCSNe to be $9.5\Ms$ and $25\Ms$, respectively.
The abundances of the other progenitors are computed by the linear interpolation of the results of our models with the progenitors of nearby $\Mms$.
Figure \ref{fig:xfe-imf} shows IMF-averaged [X/Fe] of the ejecta for cases with $m_{\rm asy}=$ 0\%, 10/3\%, 10\%, 30\%, and 50\% in panels from top to bottom.
Solid-lines with filled-squares show averaged [X/Fe] for cases with the initial abundances of 20 nuclei in a stellar evolution code~\citep{2002RvMP...74.1015W},
while dashed-lines with open-squares portray those with more than 1,300 nuclei~\citep{2002ApJ...576..323R} only for the $15.0\Ms$ and $25.0\Ms$ progenitors.
We find that IMF-averaged [X/Fe] for elements with $30 \le Z \le 40$ 
in cases with the smaller asymmetry of $m_{\rm asy} = 10/3\%$ and $10\%$ are roughly consistent with the solar abundance, although Rb and Zr are slightly underproduced and overproduced, respectively.
On the other hand, those for the elements with $30 \le Z \le 40$ with $m_{\rm asy} \ge 30\%$ are remarkably overproduced compared with the solar abundances.

It should be mentioned that the contribution to [X/Fe] for elements with $30 \le Z \le 40$ of the weak $s$-process,
which is taken into account in cases with the initial abundances of more than 1,300 nuclei (dashed-lines with open-squares), dominates over that of the $n$-rich ejecta.
We also emphasize that for the symmetric $\nu$-emission ($m_{\rm asy} = 0$; top panel) 
elements heavier than Se ($Z = 34$) are underproduced compared with the solar abundances, even if we consider the contribution of the weak $s$-process (dashed-lines with open-squares);
hence the $n$-rich ejecta synthesized via the asymmetric neutrino emissions play a complementary role to approach the solar abundance.

In short, we conclude that $\nu$ asymmetry with $m_{\rm asy} \ge 30\%$ is excluded in light of the solar abundances.
We must mention a caveat, however; $m_{\rm asy}$ for the light progenitors of $9.5\Ms$ and $11.2\Ms$ cannot be constrained from the investigation due to the small masses of the $n$-rich ejecta 
(top, left and center panels of Fig. \ref{fig:dist-ye-Mms}) and thus a small contribution to elements with $Z \ge 31$.

\begin{figure}
 \includegraphics[scale=1.1]{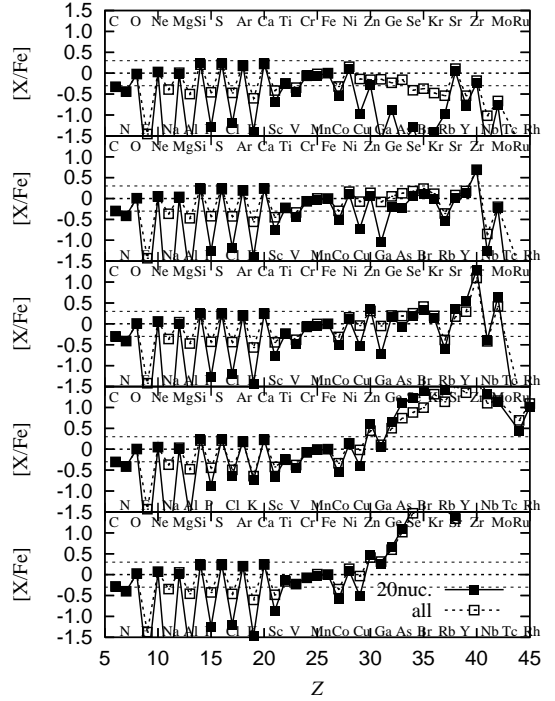}
 \vspace*{-10pt}
 \caption{
 Same as Fig.~\ref{fig:xfe-Mms} but adding the contribution from SNIa based on W7 model~\citep{2018ApJ...861..143L}. $m_{\rm asy}$ is 0\%, 10/3\%, 10\%, 30\%, and 50\% from top to bottom in each panel.
 }
 \label{fig:xfe-imf-plus-SNIa-W7}
\end{figure}

There is another caveat in the above analysis; we need to take into account the contribution from Type Ia SN (SNIa) since it is the dominant contributor to the solar abundances of the iron-group elements.
Hence, we add the contribution of SNIa to the IMF-averaged [X/Fe] of our CCSN ejecta so that [O/Fe] becomes zero.
We note that the resultant [X/Fe] with the solar abundances, which correspond to, by definition, is set to be zero following the convention in the literature.
Figure \ref{fig:xfe-imf-plus-SNIa-W7} shows [X/Fe] of the IMF-averaged CCSN ejecta added by the contribution of the SNIa ejecta of the W7 model~\citep{2018ApJ...861..143L},
in which a Ni-overproduction problem is resolved~\citep{2020ApJ...895..138K},
for cases with $m_{\rm asy}=$ 0\%, 10/3\%, 10\%, 30\%, and 50\% in panels from top to bottom.
We find that IMF-averaged [X/Fe] with the SNIa contribution well reproduce the solar abundances of elements from $Z = 6 - 39$ except for F
for cases with the smaller asymmetry of $m_{\rm asy} = 10/3\%$ and $10\%$ and cases with the initial abundances of more than 1,300 nuclei.
We note that the underproduction of F would not be a problem, since it can be complemented by the AGB stars~\citep{2018A&A...612A..16S, 2019MNRAS.490.4307O}
and rotating massive stars~\citep{2018ApJS..237...13L, 2018A&A...618A.133C} through $s$-processes.

\begin{figure*}
 \includegraphics[scale=0.65]{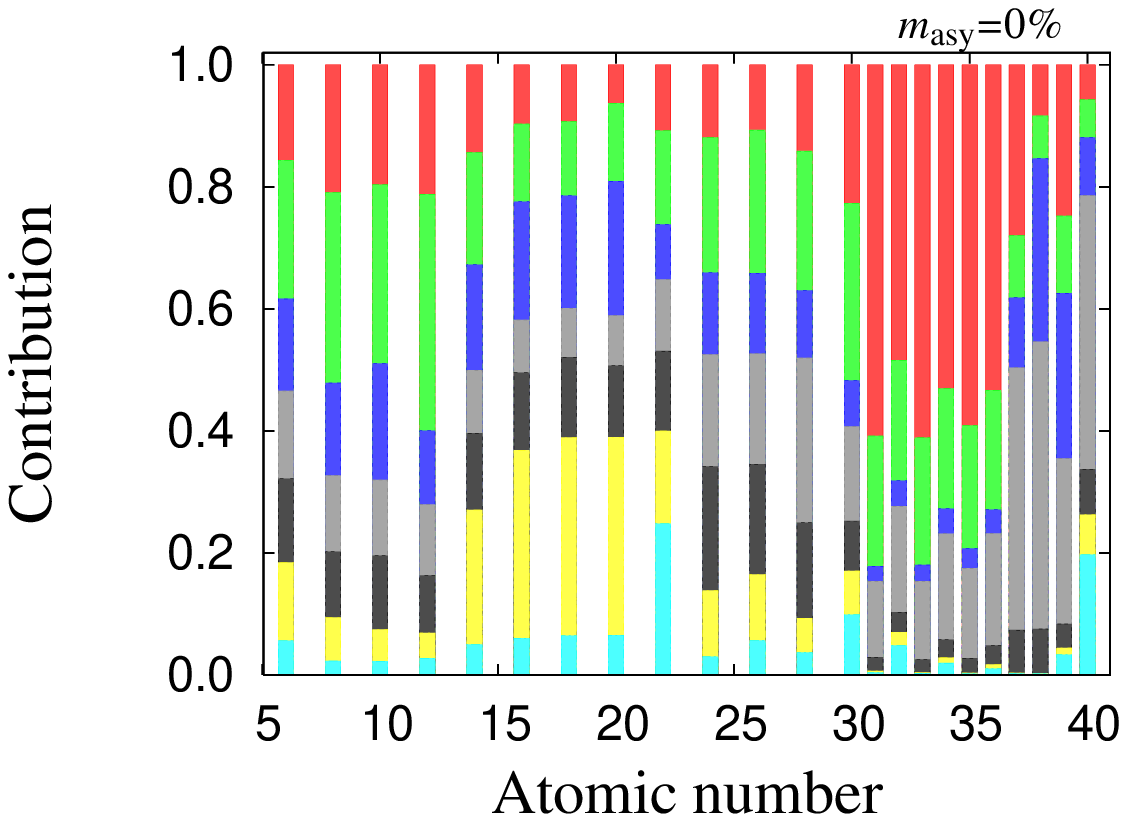}
 \includegraphics[scale=0.65]{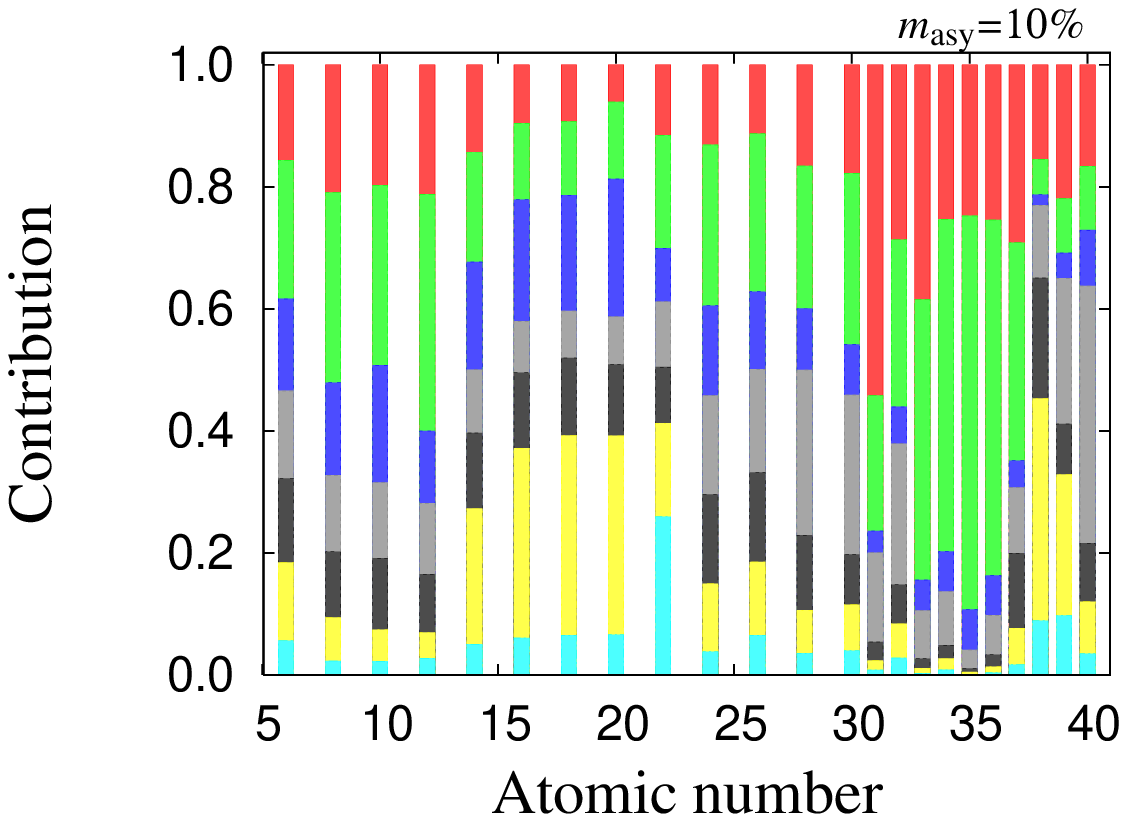}
 \vspace*{-10pt}
 \caption{
 Relative contribution to IMF-averaged [X/Fe] of the mass ranges (see text for more details) of the progenitors around $9.5\Ms, 11.2\Ms, 13.0\Ms, 15.0\Ms, 17.0\Ms, 19.4\Ms$, and $25.0\Ms$ (from bottom to top) as a function of $Z$
 for cases with $m_{\rm asy}=$ 0\% (left panel) and $m_{\rm asy}=$ 10\% (right panel).
 We omit the contribution of odd-$Z$ elements with $Z < 30$, which strongly depends on the initial compositions of each model.
 For the progenitors of $15\Ms$ and $25\Ms$, we adopt abundances of ejecta for cases with the full initial compositions as the abundances of ejecta.
 }
 \label{fig:xfe-imf-ratio}
\end{figure*}

To see the importance of each progenitor to the IMF-averaged [X/Fe] of the CCSN ejecta quantitatively,
we compute the relative contribution to the IMF-averaged [X/Fe] of the progenitors of as a function of $Z$
for cases with $m_{\rm asy}=$ 0\% and 10\% (see Figure \ref{fig:xfe-imf-ratio}).
In the computation, we adopt abundances of ejecta for cases with the full initial compositions as the abundances of ejecta for the progenitors of $15\Ms$ and $25\Ms$
and thus the contribution of $s$-process elements is partially included.
We evaluate the relative contribution of a progenitor with $M_{\rm ms}/\Ms = m_i = [9.5, 11.2, 13.0, 15.0, 17.0, 19.4, 25.0]$ for $i = 1,2, ... ,7$.
as the IMF-averaged [X/Fe] over progenitors from $M_l$ to $M_u$,
where we set $\{M_l/\Ms, M_u/\Ms\}$ to $\{(m_{i-1}+m_i)/2, (m_{i}+m_{i+1})/2\}$ for $i = 2,3, ... ,6$, $\{9.5, (m_{i}+m_{i+1})/2\}$ for $i = 1$, and $\{(m_{i-1}+m_i)/2, 25\}$ for $i = 7$.
We omit the contribution of odd-$Z$ elements with $Z < 30$ in Figure \ref{fig:xfe-imf-ratio}.
This is because the contribution is strongly inherited from the initial compositions adopted in each model
and is incorrectly dominant over that for the progenitors of $15\Ms$ and $25\Ms$ in the current evaluation.
We find that the relative contribution to elements lighter than Ni is insensitive to $m_{\rm asy}$.
On the contrary, it is very sensitive to $m_{\rm asy}$ for each element of $Z = 30-40$.
Our result suggests that the appropriate multi-D treatments are required in both CCSN and nucleosynthesis computations to determine the primary CCSN progenitor producing each element in the range of $Z = 30-40$.

\section{Asymmetric distributions of heavy elements in CCSN ejecta}
\label{ns differences}

Next, we look into the compositional differences between high-$\nue$ and $\nueb$ hemispheres,
where $p$-rich ejecta ($Y_{\rm e,1} > 0.51$) and $n$-rich ejecta ($Y_{\rm e,1} < 0.49$) abundantly exist, respectively (Fig.1 in \citet{2019MNRAS.488L.114F}).
This study would provide valuable insight into the analysis of SNR observations (see below).
Hereafter, we distinguish the two hemispheres geometrically for the case with $m_{\rm asy}= 0$ (spherical $\nu$ emission) in order to tentatively compare the cases with asymmetric neutrino emissions;
we assign $0 \le \theta \le \pi/2$ and $\pi/2 < \theta \le \pi$ as northern and southern hemispheres, respectively.

\begin{figure}
 \includegraphics[scale=0.65]{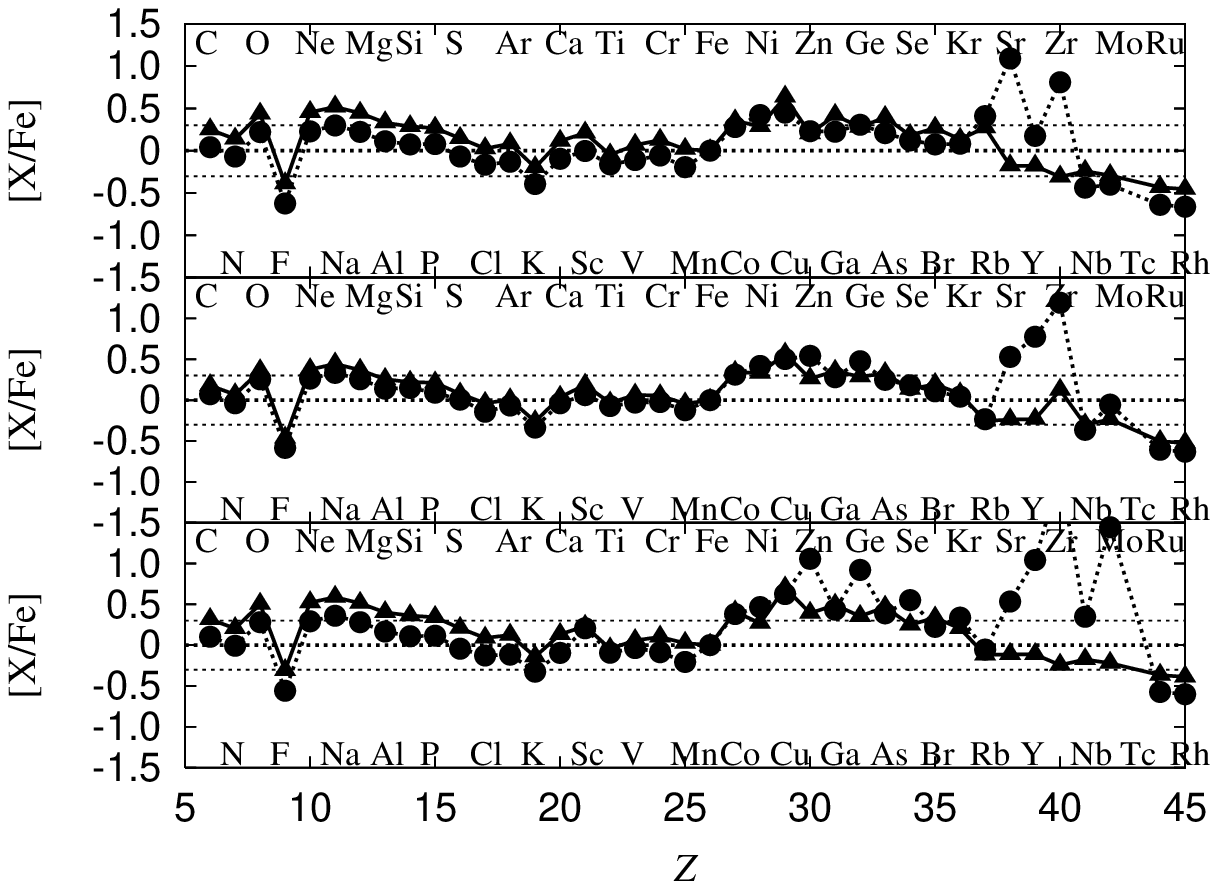}
 \includegraphics[scale=0.65]{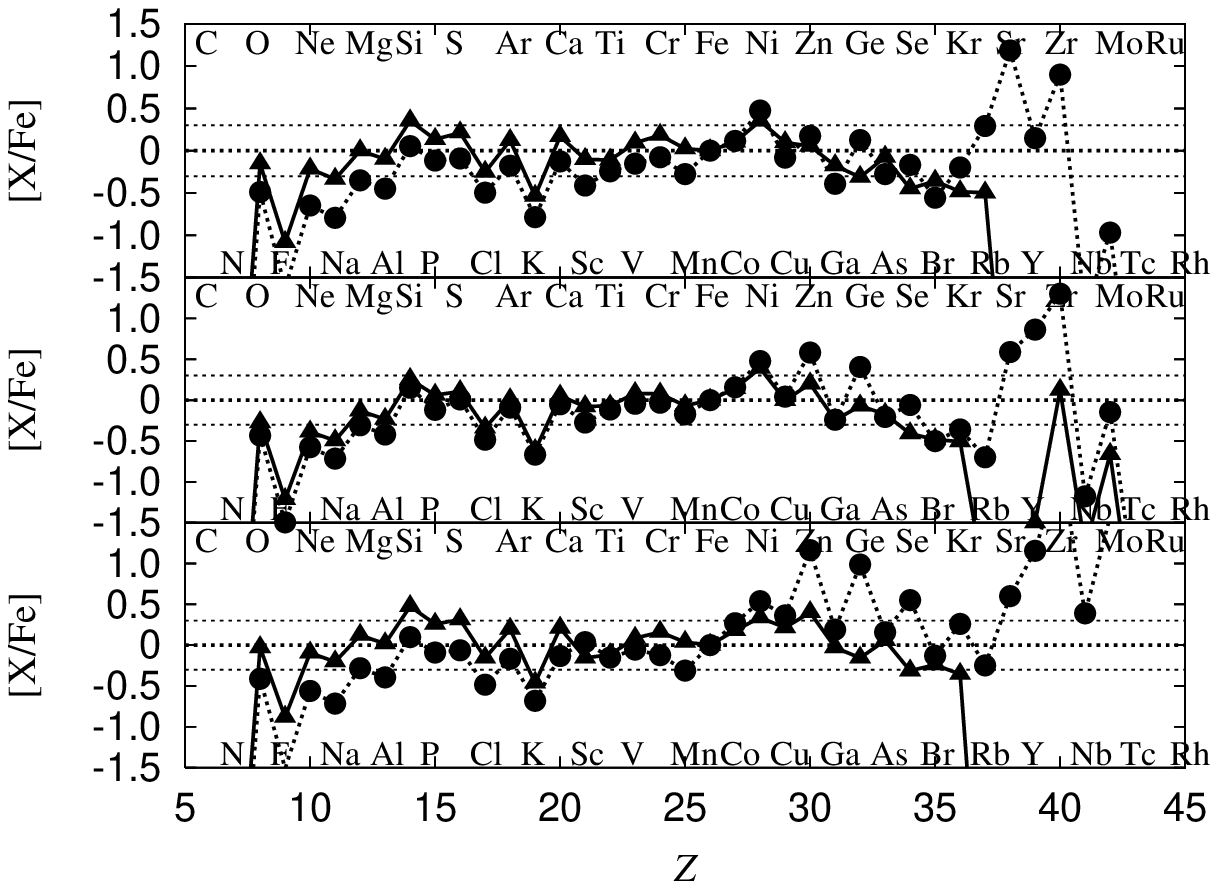}
 \caption{
 [X/Fe] of the total ejecta (top figure) and the inner one ($r_{\rm cc} \le 10,000\km$; bottom figure) for cases with $15.0\Ms$ model. 
 The solid and dotted lines denote the results  in the high-$\nue$ and high-$\nueb$ (dotted lines) hemispheres, respectively.
 $m_{\rm asy}$ is 0\%, 10/3\%, and 10\% from top to bottom in each panel.
 }
 \label{fig:xfe-ww15.0-ns}
\end{figure}

In our previous work, we showed that for the 19.4$\Ms$ progenitor, the compositional differences between the two hemispheres are small for elements lighter than Ca 
but appreciable for elements heavier than Cu (Fig.5 in \citet{2019MNRAS.488L.114F}).
The trend can be seen in the other progenitors but the 15.0$\Ms$ one is an exception.
The peculiarity of the 15.0$\Ms$ progenitor can be seen in 
Figure \ref{fig:xfe-ww15.0-ns}, which shows northern- or southern- averaged [X/Fe] of the total ejecta (top figure) and the inner ejecta (bottom figure), respectively,
for cases with $m_{\rm asy}=$ 0\% (top panels in each figure), 10/3\% (middle panels), and 10\% (bottom panels).
We note that abundances in the high-$\nueb$ hemisphere are very similar to those of all the ejecta (dotted lines with open-squares in the bottom left panel of Fig. \ref{fig:xfe-Mms})
and that the similarity is common for all progenitor models. 
On the other hand, the differences in [X/Fe] for elements lighter than Ca between high-$\nue$ (or northern) and $\nueb$ (or southern) hemispheres
emerge only in $15\Ms$ model due to the large asymmetric explosion.
The differences for Sr, Y, and Zr can be clearly seen even for the spherical $\nu$ emission (the top panels in each figure)
and those for Ni, Zn, Ge, Se, and Mo become more prominent in the cases with larger $m_{\rm asy}$.
Focusing on the inner ejecta (bottom figure), in which amounts of $s$-process elements are smaller than the total ejecta, 
we find that the differences for Ni, Zn, Ge, Se, and Mo are further enhanced.

\begin{figure}
 \includegraphics[scale=0.65]{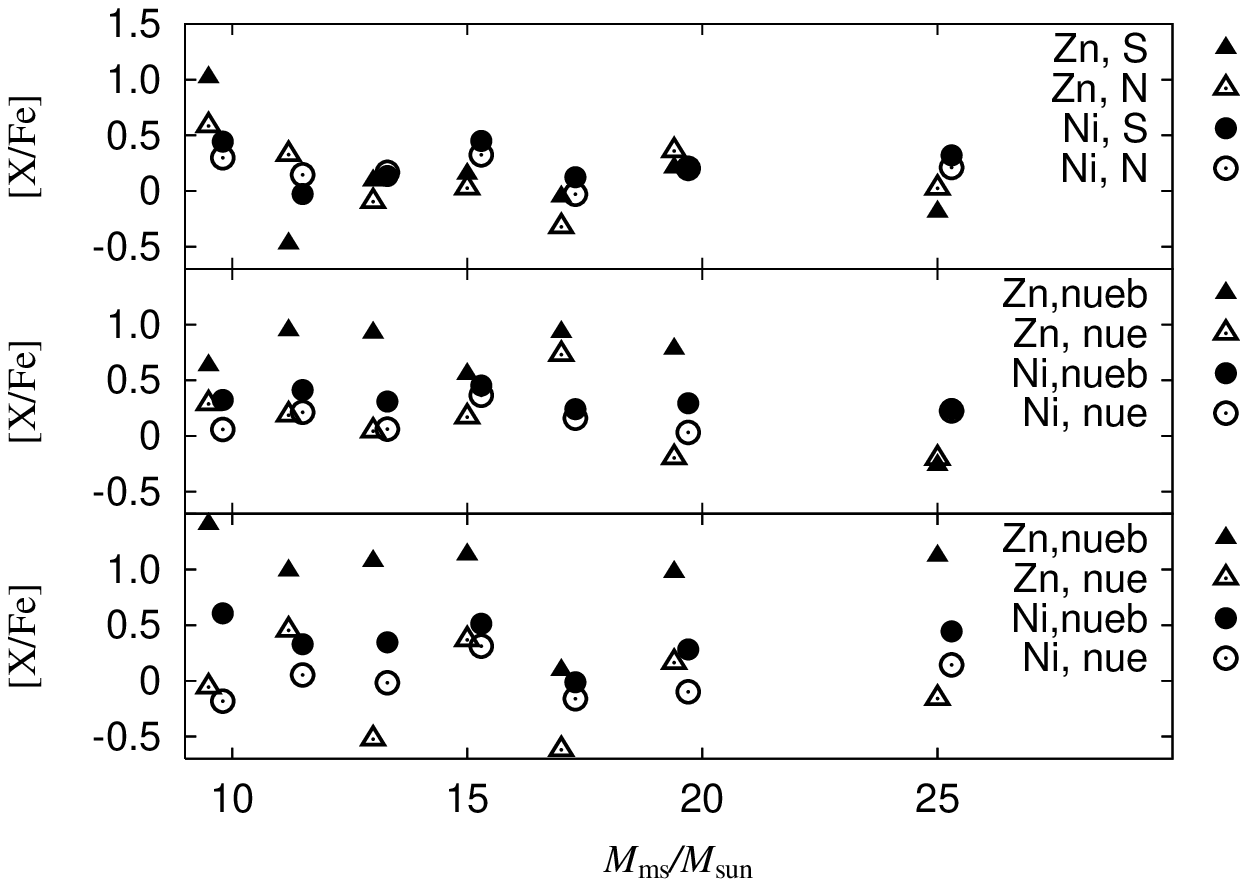}
 \caption{
 Progenitor dependence of [Ni/Fe] and [Zn/Fe] of the total ejecta.
 Filled and open circles (triangles) indicate [Ni/Fe] ([Zn/Fe]) in the high-$\nueb$ (or southern) hemispheres and the high-$\nue$ (or northern), respectively.
 We show the results with $m_{\rm asy}=$ 0\%, 10/3\%, and 10\% from top to bottom.
 For a display purpose, [Ni/Fe] (circles) slightly move to the horizontal direction.
 }
 \label{fig:xfe-ns-Ni-Zn}
\end{figure}

Figure \ref{fig:xfe-ns-Ni-Zn} shows the progenitor dependence of [Ni/Fe] and [Zn/Fe] of the ejecta in the high-$\nue$ (or northern) and the high-$\nueb$ (or southern) hemispheres
for $m_{\rm asy}=$ 0\%, 10/3\%, and 10\%.
Filled  and open circles indicate [Ni/Fe] in the high-$\nueb$ (or southern) hemispheres and the high-$\nue$ (or northern), respectively,
while filled and open triangles show [Zn/Fe] in the high-$\nueb$ (or southern) hemispheres and the high-$\nue$ (or northern), respectively.
For cases with the symmetric $\nu$ emission (top panel), the differences of both [Ni/Fe] and [Zn/Fe] between the two hemispheres are small except in [Zn/Fe] for the light progenitors of $9.5\Ms$ and $11.2\Ms$,
in which Zn is abundantly produced (top left and center panels in Fig. \ref{fig:xfe-Mms}).
The differences in these progenitors are chiefly caused by asymmetric and early-phase shock revival, which results in producing the asymmetric $n$-rich ejecta. 
We note that it is the similar case as SN explosion of a super-AGB progenitor~\citep{2011ApJ...726L..15W}.

With increasing $m_{\rm asy}$, on the other hand, the differences of [Ni/Fe] and [Zn/Fe] between the two hemispheres becomes larger regardless of progenitors, 
and they are more prominent if we remove the contribution of the O-rich ejecta.
However, We note that, although the asymmetric distribution of [Zn/Fe] is larger than that in [Ni/Fe], the ejected masses of Zn are smaller by an order of magnitude compared with those of Ni, 
indicating that the asymmetry of [Ni/Fe] may be easier to be measured than that of [Zn/Fe] in real observations (but see below). 
We also find that the asymmetric distributions of both [Ni/Fe] and [Zn/Fe] are significant for the lighter progenitors with a mass of $\le 13\Ms$.

Our result suggests that the asymmetric distributions of [Ni/Fe] and [Zn/Fe] in the ejecta would be a sign of the existence of the asymmetric neutrino emissions in a CCSN core except for very light progenitors. 
We also note that, although the mass of Zn is much smaller than Ni, the ejecta mass tends to be larger for the heavier progenitors ($\ge 15\Ms$), 
indicating that asymmetric distributions of [Zn/Fe] may be directly resolved by the future observations (see below).
For those observations, it would be crucial to spatially resolve the ejecta, since the asymmetry would be imprinted more vividly in the inner region.

\section{Conclusion}\label{sec:conclusion}

This paper is a sequel to \citet{2019MNRAS.488L.114F}, in which we investigated the impacts of the asymmetric neutrino emissions on explosive nucleosynthesis for the 19.4$\Ms$ progenitor.
In the present study, we extend the investigation to seven progenitors with a mass range from 9.5 to 25$\Ms$ to see the progenitor dependence and evaluate the IMF-averaged abundances.
We performed 35 axisymmetric, hydrodynamic simulations of the CCSN explosion in total, employed with a 2D hydrodynamic code with a simplified neutrino transport.
To measure the impacts of asymmetric neutrino emissions, we systematically change the amplitude of the dipole component of neutrino luminosities ($m_{\rm asy}$) from $0 \%$ to $50 \%$ 
(see Eqs. \ref{eq:asymmetric ne luminosity} and \ref{eq:asymmetric neb luminosity}).
The 2D hydrodynamic data is post-processed with tracer particles
and then we have evaluated abundances and masses of the SN ejecta with the nuclear reaction network of about 2,500 nuclei from $n$, $p$ to Nd.
The IMF-averaged abundances are computed with taking into account the contribution from type Ia SNe, and we find that the even-$Z$ elements lighter than Fe is less sensitive to $m_{\rm asy}$ (see Fig.\ref{fig:xfe-imf-plus-SNIa-W7}).
However, some remarkable dependence on $m_{\rm asy}$ emerges in our analyses, which are briefly summarized as follows;
\begin{enumerate}
 \item The asymmetric $\nu$ emissions lead to the abundant ejection of 
 $p$- and $n$-rich matter in the high-$\nue$ and -$\nueb$ hemispheres, respectively.
 \item The asymmetric ejection of the $n$-rich matter in the high-$\nueb$ hemisphere gives a strong impact on the abundances of elements heavier than Ni regardless of progenitors (see Fig.\ref{fig:xfe-Mms}).
 \item For $m_{\rm asy} = 10/3\%$ and $10\%$, 
       which are comparable to those appeared in recent multi-D simulations of CCSNe~\citep{2014ApJ...792...96T, 2019ApJ...880L..28N, 2019MNRAS.489.2227V}, 
       the averaged abundances for elements lighter than Y are comparable to those of the solar abundances (see Fig.\ref{fig:xfe-imf-plus-SNIa-W7}).
 \item For $m_{\rm asy} \ge 30\%$, however, the averaged abundances for elements heavier than Ge are clearly overproduced compared with the solar abundances (see Fig.\ref{fig:xfe-imf-plus-SNIa-W7}).
 \item The compositional differences of elements heavier than Ni emerge, in general, between the two hemispheres in the case with high $m_{\rm asy}$ regardless of progenitors.
       However, the detail is a bit complex; indeed, the strong asymmetric explosions (as the $15.0\Ms$ models) lead to some peculiar trends in the compositional difference (see Fig.~\ref{fig:xfe-ns-Ni-Zn}).
\end{enumerate}

Last but not least, we briefly discuss the implications of our results for future observations of SNR.
The proper motion of NS triggered by the asymmetric CCSN explosion is expected in the opposite direction to stronger shock expansion~\citep[see, e.g.,][]{2006A&A...457..963S, 2010PhRvD..82j3016N, 2012MNRAS.423.1805N, 2013A&A...552A.126W,2019ApJ...880L..28N}.
The temperature in the post-shock matter becomes higher in the stronger shock wave, which leads to the preferential synthesis of Fe and intermediate-mass elements (IME), such as Si, S, Ar and Ca, 
through the explosive Si- and O-burning, respectively.
Hence, the NS would reside in the opposite hemisphere to the one with IME-rich ejecta;
indeed, some recent observations~\citep{2017ApJ...844...84H, 2018ApJ...856...18K,2020ApJ...889..144H} suggest that there is a correlation between the flight-direction of NS and the asymmetric distributions of ejecta compositions.
We confirmed that if the asymmetric neutrino emissions exit, the $n$-rich ejecta tend to appear in the high-$\nueb$ hemisphere, 
which is the same direction of the stronger shock expansion~\citep{2014ApJ...792...96T, 2019ApJ...880L..28N, 2019MNRAS.489.2227V}, regardless of progenitors.
Therefore, the $n$-rich ejecta, which are abundant in Ni, Zn, and much heavier elements, would be located in the same direction as the IME-rich ejecta and opposed to that of the NS kick.
Our result suggests that Ni (Zn) would be preferable targets to measure the asymmetry for low (high) mass progenitors, 
although more detailed studies are called to assess the detectability quantitatively to each instrument.
This may be an urgent task since the spectroscopic X-ray observations by XRISM~\citep{2020arXiv200304962X} is around the corner.

\section*{Acknowledgements}
We are grateful to S. Katsuda for useful comments on observational aspects of Zn in SNRs. 
We also thank the anonymous referee for constructive comments that improved the content of this paper.
This work is partly supported by JSPS KAKENHI Grant Number 20K03957.

\section*{DATA AVAILABILITY}
The data underlying this article will be shared on reasonable request to the corresponding author.



\input{ms.bbl}



\appendix

\section{Neutrino-core model} \label{sec: neutrino core model} 

In our CCSN simulations, the neutrino emissions are evolved with a neutrino-core model \citep{2012ApJ...757...69U, 2016ApJ...821...38S}.
In the model, the total neutrino luminosity is evolved as, 
\begin{equation}
 L_\nu(t) = -\dot{E}_{c} +S \frac{\dot{R}_c}{R_c},
\end{equation}
where $E_c$ and $R_c$ are the total energy and radius of a proto-NS core, respectively.
$S$ denotes the surface term ~\citep[see][for the definition]{2012ApJ...757...69U}, which can be expressed as
\begin{equation}
 S = -\zeta\frac{G M_c m_{\rm acc}}{R_c},
\end{equation}
where $\zeta (>0)$ is a dimensionless parameter with the order of unity;
$M_c$ is the mass of the proto-NS core, which increases with time through the mass accretion. $m_{\rm acc}$ denotes the mass of an accretion layer enveloping the proto-NS. 
It should be mentioned that $S$ is evaluated through the gravitational energy of proto-NS, $E_g = -2GM_c^2/(5 R_c)$, in our model, by introducing a new dimensionless parameter, $f_{\rm surf}$, as
\begin{equation}
S = f_{\rm surf} E_g.
\end{equation}
This simplification allows us to evaluate $S$ without specifying $\zeta$ and $m_{\rm acc}$.

By using the virial theorem, $E_{c}$ can be expressed in terms of $E_g$ and $S$ \citep[see Eq. 1 in][]{2012ApJ...757...69U}; thus, the time derivative of $E_{c}$ can be written as,
\begin{equation}
 \dot{E}_c = \frac{3\Gamma -4}{3(\Gamma -1)} \dot{E}_g + \frac{1}{3(\Gamma -1)} \dot{S},
\end{equation}
where $\Gamma$ is the polytropic index, which is assumed to be constant in the model. 

In the present study, we have tuned the two parameters, $\Gamma$ and $f_{\rm surf}$, of the neutrino-core model 
so that the $19.4\Ms$ progenitor explodes as SN1987A-like (the explosion energy $\sim 10^{51} \rm erg$ and the ejected mass of \nuc{Ni}{56} of $(0.07-0.08) \Ms$).
The tuning results in $\Gamma = 1.7$ and $f_{\rm surf} = 0.5$ of the neutrino-core model.

Thus, the time evolution of the total neutrino luminosity can be written as
\begin{equation}
 L_\nu(t) = \frac{3\Gamma-4}{3(\Gamma-1)}(E_g +S)\frac{\dot{R_c}}{R_c}
+\frac{S -2(3\Gamma-4)E_g}{3(\Gamma-1)} \frac{\dot{M}_{\rm acc}}{M_{c}}
\end{equation}
where we assume that $\dot{M}_c$ equal to the mass accretion rate, $\dot{M}_{\rm acc}$, at the inner boundary of the computational domain in our CCSN simulations.
For the radius of the neutrino-core, we follow a model of \citet{2006A&A...457..963S}, as
\begin{equation}
R_c(t) = \frac{R_{c,b}}{1+ \left( \frac{R_{c,b}}{R_{c,f}}-1\right) \left[ 1-\exp\left( \frac{t-t_{\rm delay}}{\tau}\right) \right]},
\end{equation}
where $R_{c,b} = 80\rm km$, $R_{c,f} = 10\rm km$, $t_{\rm delay} = 50\rm \ms$, and $\tau = 1.1 \s$ are adopted in this study, which reproduce reasonably well the time evolution of $R_c(t)$ in GR1D simulations.

Having the total neutrino luminosity, we decompose it into that of each flavor. The total neutrino luminosity is the sum of those of $\nue$, $\nueb$ and $\nux$ ($\mu$, $\bar{\mu}$, $\tau$, and $\bar{\tau}$), i.e., 
\begin{equation}
  L_\nu(t) = L_{\nu_e}(t) +L_{\bar{\nu}_e}(t) +4 L_{\nu_x}(t).
\end{equation}
Following \citet{2012ApJ...757...69U, 2016ApJ...821...38S}, we assume constant fractions of the luminosities;
\begin{equation}
  L_{\nue, {\rm ave}} = f_{\nue} L_\nu(t), \,
  L_{\nueb, {\rm ave}} = f_{\nueb} L_\nu(t), \,
  L_{\nux} = f_{\nux} L_\nu(t).
\end{equation}
Here $f_{\nu_e}, f_{\bar{\nu}_e}$, and $f_{\nu_x}$ are set to be 0.25, 0.25, and 0.125, guided with the spherical simulation the core collapse of the $15.0\Ms$ progenitor.

Regarding neutrino spectra, we assume that they are all thermal with zero chemical potential. The temperatures, $T_{\nu_i}(t)$, are flavor-dependent, which are evaluated as below. 
The total neutrino luminosity can be written as
\begin{equation}
 L_\nu = \frac{7}{16} \sigma_{\rm SB} 4\pi [ T_{\nu_e}^4 R_{\nu_e}^2 + T_{\bar{\nu}_e}^4 R_{\bar{\nu}_e}^2  + 4 T_{\nu_x}^4 R_{\nu_x}^2], \label{eq:LrelationT1}
\end{equation}
where $\sigma_{\rm SB}$ is the Stephan-Boltzmann constant and $R_{\nu_i}$ denotes the neutrino sphere.
We assume that $R_c$ corresponds to the representative neutrino sphere, hence we rewrite Eq.~\ref{eq:LrelationT1} as
\begin{equation}
 L_\nu(t) = \frac{7}{16} \sigma_{\rm SB} [ T_{\nu_e}^4(t) + T_{\bar{\nu}_e}^4(t)  + 4 T_{\nu_x}^4(t) ] 4\pi R_c^2(t),
\end{equation}
where $\sigma_{\rm SB}$ is the Stephan-Boltzmann constant.
The temperature ratio among flavors is set as;
\begin{equation}
f_{T,\nu_e} : f_{T,\bar{\nu}_e} : f_{T,\nu_x} = 0.3 : 0.35 : 0.35.
\end{equation}
The ratios are determined based on the results of the spherical simulation of the core collapse of the $15.0\Ms$ progenitor, in which we refer the average energy of each species of neutrinos.
As a result, $T_{\nu_i}(t)$ can be obtained through $L_\nu(t)$ and $R_c(t)$; for instance, $T_{\nu_e}(t)$ can be obtained as
\begin{equation}
 T_{\nu_e}(t) = \left[ \frac{4 L_\nu(t)}{7\pi R_c^2(t) \sigma_{\rm SB} 
\left(
 1 +\frac{f_{E,\bar{\nu}_e}^4}{f_{E,\nu_e}^4} +4 \frac{f_{E,\nu_x}^4}{f_{E,\nu_e}^4}
\right)}
\right]^{1/4}.
\end{equation}

\section{Reversed-$L_\nu$ procedure}
\label{sec:reversed-Lnu}

As described in Sec. \ref{sec:abundances}, the direction of the $\nueb$-hemisphere in our CCSN simulations is not always the same as that with the stronger shock expansion. 
We note, however, that more elaborate multi-D CCSN simulations~\citep{2014ApJ...792...96T, 2019ApJ...880L..28N, 2019MNRAS.489.2227V} have shown that they are positively correlated with each other, 
indicating that the anti-correlation would lead to unrealistic outcomes; hence, if the anti-correlation appears, we reverse the dipole direction of neutrino emissions for nucleosynthesis computations in this study, 
albeit ad-hoc procedure.
In this appendix, we study the impact of the procedure, which is also important to understand the limitation of our results. 

\begin{figure}
 \includegraphics[scale=1.1]{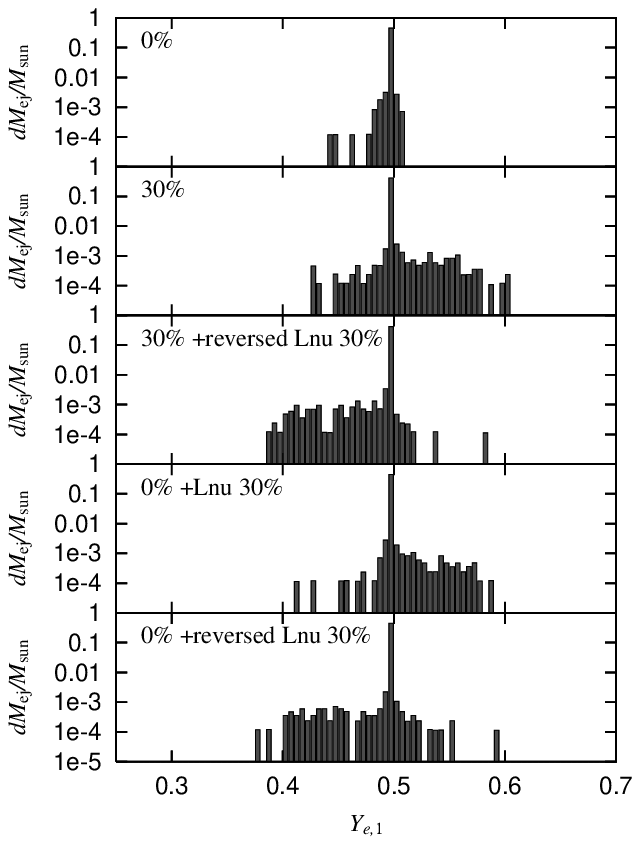}
 \vspace*{-10pt}
 \caption{
 Mass profiles of $dM_{\rm ej}$ in $Y_{\rm e,1}$ of the inner ejecta ($r_{\rm cc} \le 10,000 \km$) for the progenitor with $\Mms = 15.0\Ms$.
 Each panel corresponds to the result with a different setup of the asymmetric neutrino emissions in either CCSN simulations, nucleosynthesis computations, or both (see the text for more details).
 $m_{\rm asy}$ used in the CCSN simulations is displayed at the leftmost position in all panels.
 The middle panel corresponds to the model, in which the asymmetric neutrino emissions are reversed in the nucleosynthesis computation from those used in the CCSN simulation.
 For the fourth and fifth (bottom) panels, the asymmetric degree of the neutrino emissions is the same (zero) for the CCSN simulations (i.e., their CCSN model is identical to that shown in the top panel),
 meanwhile, we set the asymmetric degree as 30\% in the nucleosynthetic computations, and the dipole direction is the opposite between the two models.
 }
 \label{fig:dist-ye-ww15.0-asym0.3}
\end{figure}

Figure \ref{fig:dist-ye-ww15.0-asym0.3} shows mass profiles of $dM_{\rm ej}$ in $Y_{\rm e,1}$ of the ejecta from the inner region ($r_{\rm cc} \le 10,000\km$) for the progenitor with $\Mms = 15.0 \Ms$.
The top and second panels show the mass profiles for cases with $m_{\rm asy}=0\%$ and $m_{\rm asy}=30\%$, respectively,
in which the direction of $\nu$-asymmetry for the hydrodynamic simulations is set to be the same as that in the nucleosynthetic calculation.
The direction of the $\nueb$-hemisphere for the case with $m_{\rm asy}=30\%$, however, does not correlate with that of stronger shock.
The third panel displays the result with the reversed dipole direction in the nucleosynthesis computation to recover the positive correlation between the $\nueb$-hemisphere and that of the stronger shock expansion.
By comparing the second and third panels, we find that the strong $p$-rich ejecta in the former disappears in the latter, but more $n$-rich ejecta appears, instead.
To understand the trend more clearly,
we study another case in which the asymmetric degree and dipole direction are the same as those used in the case with the second panel, 
but we employ a CCSN model with $m_{\rm asy}=0\%$ as a background; the result is displayed in the fourth panel.
We confirm that the strong $p$-rich ejecta is observed in both the second and fourth panels,
indicating that it is an artifact due to the negative correlation between the $\nueb$-hemisphere and that of the stronger shock expansion.
In other words, the $\nue$-hemisphere becomes the same direction as that of the stronger shock expansion, implying that the $\nue$ absorption is artificially enhanced and then $Y_e$ of the ejecta is increased accordingly.
To strengthen this argument, we study another model, in which we carry out the nucleosynthesis computation with the reversed neutrino asymmetry with $m_{\rm asy}=30\%$ under the CCSN simulation with $m_{\rm asy}=0\%$;
the result is displayed in the fifth (bottom) panel.
As seen in the panel, the $p$-rich ejecta is reduced, and the mass profile approaches that of the third panel.
Our results suggest that the dipole direction of the neutrino emissions in nucleosynthesis computations is a more influential parameter than that in the CCSN simulation.
Hence, we conclude that the adjustment of the dipole direction to hold the positive correlation between the $\nueb$-hemisphere and that of the stronger shock expansion in the nucleosynthesis computation 
is appropriate to capture the qualitative trend of the asymmetry of ejecta composition.
We also confirm through this study that the uncertainty by the procedure is the same level as the differences between the third and bottom panels.


\bsp	
\label{lastpage}
\end{document}